\documentclass[sn-mathphys]{sn-jnl}

\jyear{2021}%

\theoremstyle{thmstyleone}%
\theoremstyle{thmstyletwo}%

\theoremstyle{thmstylethree}%

\begin{document}

\title[\ ]{Gyrokinetic theory of toroidal Alfv\'en eigenmode saturation via nonlinear wave-wave coupling}


\author[1,2]{\fnm{Zhiyong} \sur{Qiu}}\email{zqiu@zju.edu.cn}

\author[1,2,3]{\fnm{Liu} \sur{Chen}}\email{liuchen@zju.edu.cn}

\author[1,2]{\fnm{Fulvio} \sur{Zonca}}\email{fulvio.zonca@enea.it}

\affil[1]{\orgdiv{Institute for Fusion Theory and Simulation, School of Physics}, \orgname{Zhejiang University}, \orgaddress{  \city{Hangzhou}, \postcode{310027}, \country{China}}}

\affil[2]{\orgdiv{Center for Nonlinear Plasma Science and  C. R.  ENEA Frascati},  \orgaddress{\postcode {C. P. 65, 00044}, \city{Frascati},   \country{Italy}}}

\affil[3]{\orgdiv{Department of   Physics and Astronomt}, \orgname{University of California}, \orgaddress{\city{Irvine}, \postcode{92697-4575}, \state{CA}, \country{U.S.A.}}}

\abstract{Nonlinear wave-wave coupling constitutes an important route for the turbulence spectrum evolution in both space and laboratory plasmas. For example, in a reactor relevant fusion plasma, a rich spectrum of symmetry breaking shear Alfv\'en wave (SAW) instabilities are expected to be excited by energetic fusion alpha particles, and   self-consistently determine the anomalous alpha particle transport rate by the saturated electromagnetic perturbations. In this work, we will show that the nonlinear gyrokinetic theory is a  necessary and powerful tool in qualitatively and quantitatively investigating the nonlinear wave-wave coupling processes. More specifically, one needs to employ the gyrokinetic approach in order  to account for the breaking of the ``pure Alfv\'enic state" in the short wavelength kinetic regime, due to the short wavelength structures associated with nonuniformity intrinsic to magnetically confined plasmas.\\

Using well-known toroidal Alfv\'en eigenmode (TAE) as a paradigm case, three nonlinear wave-wave coupling channels expected to significantly influence the TAE nonlinear dynamics are investigated to demonstrate the strength and necessity of nonlinear gyrokinetic theory in predicting crucial processes in a future reactor burning plasma. These are: 1. the nonlinear excitation of meso-scale zonal field structures via modulational instability and TAE scattering into short-wavelength stable domain; 2. the TAE frequency cascading due to nonlinear ion induced scattering and the resulting saturated TAE spectrum; and 3. the cross-scale coupling of TAE with micro-scale ambient drift wave turbulence and its effect on TAE regulation and anomalous electron heating.
}



\maketitle


\section{Introduction}

Shear Alfv\'en waves (SAWs) \cite{HAlfvenNature1942} are fundamental  electromagnetic fluctuations  in magnetised plasmas,  and are ubiquitous in space and laboratories.    SAWs exist due to  the balance of restoring force due to magnetic  field line bending and   plasma  inertia, and  
 are characterized by transverse magnetic perturbations propagating along equilibrium magnetic field lines,   with the parallel wavelength comparable to system size, while perpendicular wavelength varying from system size to ion Larmor radius. Due to their incompressible character, SAWs can be driven unstable with a lower threshold  in comparison to that of compressional Alfv\'en waves or ion acoustic waves.           In magnetically confined plasmas typical of  fusion reactors such as ITER \cite{KTomabechiNF1991} and CFETR \cite{YWanNF2017}, with their phase/group velocity comparable to the characteristic speed  of super-thermal fusion alpha particles,  SAW
instabilities could be strongly excited by fusion alpha particles as well as energetic particles (EPs) from auxiliary heating. The enhanced symmetry-breaking SAW fluctuations could lead to   transport loss of   EPs across magnetic field surfaces; raising   an important challenge to the good EP confinement required for sustained burning \cite{AFasoliNF2007,LChenRMP2016}.

In magnetic confined fusion devices, due to the nonuniformities associated with equilibrium magnetic geometry and plasma profile, SAW  frequency varies continuously across the magnetic surfaces and forms a continuous spectrum \cite{HGradPT1969}, on which SAWs suffer continuum damping by mode conversion to small scale structures Landau damped, predominantly, by electrons \cite{LChenPoF1974,AHasegawaPoF1976,LChenRMPP2021}. As a result, SAW instabilities can be excited as various kinds of  EP continuum modes (EPMs)  when the EP resonant  drive overcomes continuum damping \cite{LChenPoP1994},  or as discretised Alfv\'en eigenmodes (AEs) inside continuum  gaps to minimise the continuum damping, among which, the famous toroidal Alfv\'en eigenmode (TAE) \cite{CZChengAP1985,LChenVarenna1988,GFuPoFB1989,KWongPRL1991} is a celebrated example.  For a thorough understanding of the SAW instability spectrum in reactors, interested readers may refer to Refs. \cite{GVladRNC1999,AFasoliNF2007,FZoncaPoP2014b,LChenRMP2016,YTodoRMPP2019} for comprehensive reviews.

The SAW instability induced EP anomalous  transport/acceleration/heating  rate,  depends on the SAW instability amplitude and spectrum via wave-particle resonance conditions \cite{LChenJGR1999,LChenJGR1991}, which are, determined by the nonlinear saturation mechanisms. The first channel for SAW instability nonlinear saturation is the nonlinear wave-particle interactions, i.e., the acceleration/deceleration of EPs  by SAW instability induced EP ``equilibrium" distribution function evolution and the consequent self-consistent SAW spectrum evolution, among which there are well-known and broadly used models introduced by Berk et al \cite{HBerkPoFB1990a,HBerkPoFB1990b,HBerkPoFB1990c}  by analogy  to the wave-particle trapping in one-dimensional beam-plasma instability system \cite{TOneilPoF1965}. More recently, Zonca et al systematically developed the non-adiabatic wave-particle interaction theory, based on nonlinear evolution of  phase space zonal structures (PSZS) \cite{FZoncaJPCS2021,FZoncaNJP2015,MFalessiPoP2019,MFalessiNJP2023,FZoncaPPCF2015}, i.e., the phase space structures that are un-damped by collisionless processes.  The PSZS approach, by definition of the ``renormalised"  nonlinear equilibria typically varying on the mesoscales  in the presence of microscopic turbulences,   self-consistently describes the EP phase space non-adiabatic evolution and  nonlinear evolution of turbulence due to varying EP ``equilibrium" distribution function, very often in the form of non-adiabatic frequency chirping, and is described by a closed Dyson-Schr$\ddot{o}$dinger model \cite{FZoncaJPCS2021}.  Both processes  are tested and corresponding theoretical frameworks are broadly used   in interpretation of experimental results as well as large scale numerical simulations, e.g., \cite{XWangPRE2012,HZhangPRL2012,LYuEPL2022}.  The other channel for SAW nonlinear evolution, relatively less explored in large-scale simulations, is the nonlinear wave-wave coupling mechanism, describing SAW instability spectrum evolution due to interaction with other electromagnetic oscillations, and is the focus of the present brief review using TAE as a paradigm case. These approaches developed for TAE and the obtained results, are general, and can be applied to other SAW instabilities based on the knowledge of their linear properties.

The nonlinear wave-wave coupling process, as an important route for SAW instability nonlinear dynamic evolution and saturation \cite{LChenPoP2013}, is expected to be even   more important in burning plasmas of future reactors;   where, different from present-day existing magnetically confined devices, the EP power density can be comparable with that of bulk thermal plasmas, and the EP characteristic orbit size is much smaller than the system size. As a consequence, there is a rich spectrum of SAW instabilities in future reactors \cite{AFasoliNF2007,LChenRMP2016,TWangPoP2018,ZRenNF2020}, with most unstable modes being characterized by $n\gtrsim O(10)$ for maximized wave-particle power exchange,    with $n$ being the toroidal mode number. That is, multi-$n$ modes with comparable  linear growth rates could be excited simultaneously.  These SAW instabilities are,  thus, expected to interact with each other, leading to complex spectrum evolution that eventually affects the EP transport. It is noteworthy that  the nonlinear wave-particle   interaction, described by Dyson Schr$\ddot{o}$dinger model and nonlinear wave-wave couplings embedded within a generalized nonlinear Schr$\ddot{o}$dinger equation, are two pillars of the unified theoretical framework for self-consistent  SAW nonlinear evolution and EP transport, as summarized in Ref. \cite{FZoncaJPCS2021},  which is being actively developed by the Center for Nonlinear Plasma Physics (CNPS) collaboration \footnote{For more information and activities of CNPS, one may refer to the CNPS homepage at https://www.afs.enea.it/zonca/CNPS/}.

Due to the typically short scale structures associated with continuous spectrum, the nonlinear couplings of   SAW instabilities   are dominated by the perpendicular nonlinear scattering via  Reynolds and Maxwell stresses, instead of the polarization nonlinearity \cite{AHasegawaPoF1976,LChenEPL2011,RSagdeevbook1969}. Thus, the kinetic treatment is needed to capture the essential ingredients of SAW nonlinear wave-wave coupling dominated by small structures that naturally  occur due to SAW continuum, and some other fundamental physics not included in magnetohydrodynamic (MHD) theory, e.g., the wave-particle interaction crucial for ion induced scattering of  TAEs \cite{TSHahmPRL1995,ZQiuNF2019a}, and trapped particle effects  in the low frequency range that may lead to neoclassical inertial enhancement, which plays a key role   for zonal field structure (ZFS) generation \cite{MRosenbluthPRL1998,LChenPoP2000,LChenPRL2012}.  These crucial physics ingredients  are not included in the MHD description, and kinetic treatment is mandatory to both quantitatively and qualitatively study the nonlinear wave-wave coupling processes of SAWs. These features can be fully and conveniently covered by nonlinear gyrokinetic theory \cite{EFriemanPoF1982} derived by systematic removal of fast gyro motions with $\Omega_{c}\gg\omega_{A}$, and yield quantitatively, using TAE as a paradigm case, the nonlinear saturation level and corresponding EP transport and/or heating. The general knowledge obtained here, as noted in the context of this review, can be straightforwardly applied to other kinds of SAW instabilities, with the knowledge of  their linear properties.

The rest of the paper is organized as follows. In Sec. \ref{sec:SAW_nl}, the general background knowledge of nonlinear wave-wave coupling of SAW instabilities in toroidal geometry are introduced, where SAW instabilities in toroidal plasmas and nonlinear wave-wave coupling are briefly reviewed.   The kinetic theories of TAE saturation via nonlinear wave-wave coupling are reviewed in Sec. \ref{sec:TAE_nl}, where three channels for TAE nonlinear dynamic evolution are introduced. Finally, a brief summary is given in Sec. \ref{sec:summary}.

\section{Theoretical framework of nonlinear mode coupling and SAWs in toroidal plasmas}
\label{sec:SAW_nl}

In this section, the basic elements  needed for SAW nonlinear mode coupling are introduced, including the linear SAW dispersion relation,  pure Alfv\'enic state, perpendicular nonlinear coupling,  and nonlinear gyrokinetic theoretical framework.  For   accessibility to general readers, these materials are introduced in a pedagogical way.  Readers interested in more technical details may refer  to references given.

\subsection{Nonlinear wave-wave coupling}

The nonlinear wave-wave coupling corresponds to wave spectrum evolution due to  interaction with other collective oscillations, and is an important pillar of nonlinear plasma physics \cite{RSagdeevbook1969}. For SAW instability, there is an important property that, in uniform plasmas and ideal MHD limit,  the Reynolds and Maxwell stresses, will exactly cancel each other. Thus, SAWs can grow  to large amplitudes without being distorted by nonlinear effects. This is called ``pure Alfv\'enic state",  and will be addressed briefly below. As a result, for the nonlinear mode couplings of SAWs, the pure Alfv\'enic state shall be broken by, e.g., system nonuniformity and/or kinetic compression, as addressed in Ref.  \cite{LChenPoP2013}.

The momentum equation for the incompressible SAW  nonlinear  evolution in the low $\beta$ plasma limit, keeping up to quadratic terms, can be written as
\begin{eqnarray}
\rho_0(\partial_t+\delta\mathbf{v}\cdot\nabla)\delta\mathbf{v}= \delta\mathbf{J}\times \mathbf{B}_0/c+\delta\mathbf{J}\times\delta\mathbf{B}/c,\label{eq:momentum_MHD}
\end{eqnarray}
with $\rho$ being the mass density, $\mathbf{v}$  the fluid  velocity,  $\mathbf{J}$  the current density,   $\mathbf{B}$   the magnetic field,  and $\delta$ indicating perturbed quantities. 
Equation (\ref{eq:momentum_MHD}),  together with the Ampere's law
\begin{eqnarray}
\nabla\times \delta \mathbf{B}=4\pi\delta \mathbf{J}/c 
\end{eqnarray}
and the Faraday's law with ideal MHD condition embedded,   
\begin{eqnarray}
\partial_t\delta\mathbf{B} =\nabla\times(\delta\mathbf{v}\times\mathbf{B}_0),\label{eq:faraday}
\end{eqnarray}
yield, in the linear limit, 
\begin{eqnarray}
\frac{\delta\mathbf{v}}{V_A}=\pm\frac{\delta \mathbf{B}}{B_0},\label{eq:walen_relation}
\end{eqnarray}
which correspond  to the famous Walen relation \cite{CKierasJPP1982}. In deriving equation (\ref{eq:walen_relation}), the linear SAW dispersion relation, derived from linearised  equations (\ref{eq:momentum_MHD}) and (\ref{eq:faraday}), $\omega^2=k^2_{\parallel}V^2_A$ is used, with $V_A\equiv\sqrt{B^2_0/(4\pi\rho_0)}$ being the Alfv\'en velocity.

Equation (\ref{eq:momentum_MHD}), in the nonlinear limit,  can be re-written as
\begin{eqnarray}
\rho_0\partial_t\delta\mathbf{v}^{(2)}= -\nabla \lvert\delta B\rvert^2/(8\pi)-\mbox{MX}-\mbox{RS},
\end{eqnarray}
with $\mbox{MX}\simeq-\delta\mathbf{B}_{\perp}\cdot\nabla\delta \mathbf{B}_{\perp}/(4\pi)$ and $\mbox{RS}\equiv \rho_0\delta\mathbf{v}_{\perp}\cdot\nabla\delta\mathbf{v}_{\perp}$ being, respectively, the   Maxwell and Reynolds stresses, and the first term on the right hand side corresponding to the parallel ponderomotive force \cite{RSagdeevbook1969}, which is typically much smaller than RS and MX due to the typical $k_{\parallel}\ll k_{\perp}$ ordering.  It can be seen clearly that, in the present model of ideal MHD, uniform plasma limit, RS and MX  cancel each other, so SAW can grow  to large amplitude without being distorted by nonlinear processes. Thus, to understand the nonlinear evolution of SAW instabilities as   this pure Alfv\'enic state is broken, higher order nonlinearities that occur on longer time  scales should be  introduced; i.e., it is necessary to go beyond    the ideal MHD description. As we shall show in the following applications using TAE as an example, plasma nonuniformity as well as plasma compressibility may play crucial roles in breaking the Alfv\'enic state for different control parameters.     In order to account for these effects  for SAWs as well as drift waves (DWs) involved in the analysis with frequencies much lower than ion cyclotron frequency, nonlinear gyrokinetic theory is shown to be extremely useful in studying the nonlinear wave-wave interaction physics, and is introduced in the Sec. \ref{sec:theoretical_model}.   For a thorough discussion of pure Alfv\'enic state and SAW/KAW nonlinear dynamics as it  is broken by various effects, interested readers may refer to Ref. \cite{LChenPoP2013} for more details.  

\subsection{Nonlinear gyrokinetic theoretical framework}\label{sec:theoretical_model}

The nonlinear gyrokinetic equation is derived by systematic  removal of the fast gyro-motion of particles, noting the conservation of magnetic moment $\mu\equiv m v^2_{\perp}/(2B)$ in the low frequency regime with $\omega\ll\Omega_{c}$, and it is a powerful  tool in theoretical/numerical studies  of low frequency fluctuations of interest in magetically confined plasmas \cite{EFriemanPoF1982,ABrizardRMP2007,HSugamaRMPP2017}. In gyrokinetic theory, the fluctuating particle response can be separated into adiabatic and non-adiabatic components, 
\begin{eqnarray}
\delta f_j=\left(\frac{q}{m}\right)_j\delta\phi_k \frac{\partial}{\partial E} F_{0j} +\exp(-\mathbf{\rho}\cdot\nabla) \delta H_j,
\end{eqnarray}
with the non-adiabatic particle response derived from nonlinear gyrokinetic equation \cite{EFriemanPoF1982}
\begin{eqnarray}
&&\left(\partial_t+v_{\parallel}\mathbf{b}\cdot\nabla + \mathbf{v}_d\cdot\nabla\right)\delta H_k \nonumber\\
&=& i\frac{q}{m}\left(\omega\partial_E +\hat{\omega}_*\right) F_{0} J_k\delta L_k - \sum_{\mathbf{k}=\mathbf{k}'+\mathbf{k}''} \Lambda^{k'}_{k''} J_{k'}\delta L_{k'}\delta H_{k''}.\label{eq:gke}
\end{eqnarray}
Here, $E = v^2/2$ is the energy per unit mass, $\mathbf{v}_d=\mathbf{b}\times[(v^2_{\perp}/2)\nabla\ln B_0+v^2_{\parallel}\mathbf{b}\cdot\nabla\mathbf{b}]$ is the magnetic drift,   $\hat{\omega}_* \equiv \mathbf{k}\cdot\mathbf{b}\times \nabla \ln{F_0}/\Omega_c$ is  related to the diamagnetic drift frequency associated with plasma nonuniformities.
 In the present work focusing on the nonlinear evolution of TAE with prescribed amplitude due to nonlinear mode coupling, with  dominant role played by thermal plasma contribution to RS  and MX \footnote{We incidentally note that  EP may contribute significantly to ZFS generation by TAE as TAEs are exponentially growing due to wave-particle resonance, and lead to the ``forced driven" excitation of ZFS by TAE \cite{YTodoNF2010,ZQiuPoP2016}. We, however, will not discuss this case in the present review aiming at giving a fundamental picture of  TAE nonlinear dynamics via nonlinear mode coupling. }, in the rest of the manuscript, Maxwellian distribution function  is adopted for   thermal plasmas, and one has $\partial_E F_{M}=-(m/T) F_{M}$,  $\hat{\omega}_* F_M=(m/T)\omega_*F_M$ with $\omega_*\equiv -i(cT/qB_0)_j \mathbf{b}\times\nabla\ln F_M\cdot\nabla =ck_{\theta}T_j/(qB_0L_n)_j \left[1+\eta (E/T-3/2)\right]_j$, $k_{\theta}$ is the poloidal wavenumber,  and  $\eta=L_n/L_T$ with $L_n$ and $L_T$ being respectively the characteristic scale length of density and temperature nonuniformities. Furthermore,   $J_k\equiv J_0(k_{\perp}\rho)$ is the Bessel function of zero-index accounting for finite Larmor radius effects, $\delta L_k\equiv \delta\phi-v_{\parallel}\delta A_{\parallel}/c$, and $\Lambda^{k'}_{k''}\equiv (c/B_0)\mathbf{b}\cdot\mathbf{k''}\times\mathbf{k'}$ accounts for perpendicular scattering with the constraint on  wavenumber   matching condition given by $\mathbf{k}=\mathbf{k}'+\mathbf{k}''$.  In the rest of the paper, $-i\partial_l \delta\psi_k=\omega\delta A_{\parallel k}/c$ is introduced for conveniently treating the inductive parallel electric field component, which allows recovering the ideal MHD condition ($\delta E_{\parallel}=0$)   by straightforwardly taking $\delta\psi_k=\delta\phi_k$.

The governing equations are derived from quasi-neutrality condition
\begin{eqnarray}
\frac{n_0e^2}{T_i}\left(1+\frac{T_e}{T_i}\right)\delta\phi_k=\sum_{j=e,i} \left\langle q J_k\delta H_k\right\rangle,\label{eq:qn}
\end{eqnarray}
with angular brackets denoting velocity space integration and,   with magnetic compression being negligible in the low-$\beta$ limit of interest here,  the   nonlinear gyrokinetic vorticity equation becomes \cite{LChenJGR1991,LChenNF2001}
\begin{eqnarray}
&&\frac{c^2}{4\pi\omega^2} B\frac{\partial}{\partial l}\frac{k^2_{\perp}}{B}\frac{\partial}{\partial l}\delta\psi_k + \frac{e^2}{T_i}  \left(1-\frac{\omega_*}{\omega}\right)_k  \left\langle\left(1-J^2_k\right)F_M\right\rangle\delta\phi_k   -\sum_{j=e,i}\left\langle qJ_0\frac{\omega_d}{\omega}\delta H\right\rangle_k\nonumber\\
&=& -\frac{i}{\omega_k} \sum_{\mathbf{k}=\mathbf{k}'+\mathbf{k}''} \Lambda^{k'}_{k''} \left[\left\langle e(J_kJ_{k'}-J_{k''})\delta L_{k'}\delta H_{k''}\right\rangle +\frac{c^2}{4\pi}k''^2_{\perp}  \frac{\partial_l\delta\psi_{k'}\partial_l\delta\psi_{k''}}{\omega_{k'}\omega_{k''}}\right].
\label{eq:vorticity}
\end{eqnarray}
Nonlinear gyrokinetic vorticity equation is derived from parallel Ampere's law, quasi-neutrality condition and nonlinear gyrokinetic equation, and it forms, together with quasi-neutrality condition, equation (\ref{eq:qn}),   a closed set of equations describing the dynamics of low frequency fluctuations in low $\beta$ plasmas.  Note that,  for the application in the present review, in equation (\ref{eq:vorticity}), only effects associated with plasma density nonuniformity are accounted for,  while effects associated with temperature gradients are neglected systematically, i.e., $\eta\equiv L_n/L_T=0$.
The terms on the  left hand side of equation (\ref{eq:vorticity}) are, respectively, the field line bending, inertia and curvature-pressure coupling terms. This  clearly shows  the convenience of the present formulation based on the gyrokinetic  vorticity equation in studying SAW related physics, since field bending and inertia terms balance at the leading order for these fluctuations. The terms on the right hand side, on the other hand, are the formally nonlinear generalized gyrokinetic RS and MX, dominated, respectively,  by ion and electron contributions.    

In this brief review focusing on the   TAE physics due to nonlinear wave-wave interactions, TAE with prescribed amplitude are assumed, while  EPs contribution is typically small. Thus,  we include only the thermal plasma contribution in the above governing  equations.  The EPs, crucial for the TAE excitation,  can also  be important in ZFS generation during the TAE   exponential  growth phase due to resonant EP drive.   This interesting topic of nonlinear ZFS  forced driven process connected with the nonlinear EP response is beyond the scope  of the present review  and will only be  briefly discussed in  Sec.  \ref{sec:ZFS_excitation}. 

Note that, for TAE of interest of the present review, with frequency typically much larger than thermal plasma diamagnetic frequency, the system nonuniformity associated with $\omega_*$ is typically weak and, thus,  systematically neglected in the majority of  present review on TAE nonlinear physics. It is maintained, however, in Sec. \ref{sec:TAE_eDW_scattering}  in the analysis of  TAE scattering by DWs,  where finite  $\omega_*$ is crucial for the high-n DW physics, as well as  for the enhancement of nonlinear scattering rates due to the $\lvert\omega_*\rvert\gg \lvert v_i/(qR_0)\rvert$ ordering. 

\subsection{SAW instabilities in toroidal plasmas}
\label{sec:saw_torus}

In this section, the  SAW dispersion relation in the WKB limit will be derived, which is then used to symbolically  demonstrate the formation of   SAW continuum structure and the existence of discrete Alfv\'en eigenmode, using the well-known TAE as an example. The obtained linear particle responses  can be used in the following analysis of TAE nonlinear dynamics via nonlinear wave-wave coupling processes.  For the convenience of following analysis on nonlinear wave-wave couplings, the particle responses to SAW, are derived in real space, and the obtained mode equation, will be solved by transforming into  ballooning space.  Note that for TAE of interest here, $\lvert\omega_*/\omega\rvert\ll1$ is satisfied for most unstable TAEs with perpendicular wavelength  comparable to EP drift orbit width; so that the thermal plasma $\omega_*$ effects   on SAW dispersion relation are expected to be small. Thus, in the majority of the paper, the $\omega_*$ effects on TAE/KAW dispersion relation are systematically neglected.
  However, in our derivation of linear thermal plasma response to SAW  $\omega_*$ correction is kept, which will be used in Sec. \ref{sec:TAE_eDW_scattering},  where $\omega_*$ effects on KAW can be important  due to its relatively high toroidal mode number due to momentum conservation in high-n DW scattering.

\begin{figure}
\includegraphics[scale=0.55]{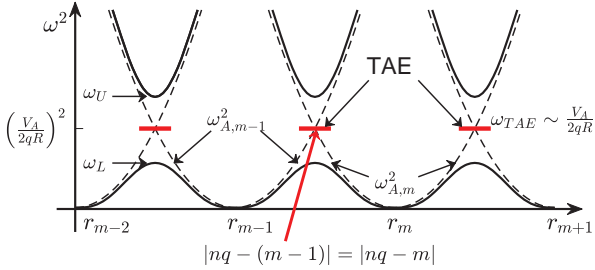} 
\caption{Toroidicity induced SAW continuum gap. The horizontal axis is radial position with $r_m$ denoting the $q=m/n$ rational surface, and vertical axis corresponds to $\omega^2$. The dashed and solid curves correspond to  the SAW continuum in the cylindrical   and toroidal limits, respectively; and $\omega_U$ and $\omega_L$ denote the upper and lower accumulational points of toroidicity induced continuum gap. }\label{fig:TAE_gap}
\end{figure}

The linear electron response to SAW  can be derived noting the $\lvert\omega/k_{\parallel} v_e\rvert\ll 1$ ordering, and one obtains
\begin{eqnarray}
\delta H_{ke}\simeq -\frac{e}{T_e} F_M\left(1-\frac{\omega_{*e}}{\omega}\right)_k\delta\psi_k. \label{eq:linear_e_TAE}
\end{eqnarray}
Meanwhile, assuming unity charge for simplicity in calculating the ion response, and noting  the $\lvert\omega\rvert\gg \lvert k_{\parallel}v_i\rvert \gg\lvert\omega_d\rvert$ ordering,  one has, at the leading order,
\begin{eqnarray}
\delta H^{(0)}_{ki}\simeq \frac{e}{T_i} F_M J_k\left(1-\frac{\omega_{*i}}{\omega}\right)_k\delta\phi^{(0)}_k.\label{eq:linear_i_TAE}
\end{eqnarray} 
Substituting into quasi-neutrality condition, one obtains,
\begin{eqnarray}
\delta\psi^{(0)}_k =  \sigma_{*k}\delta\phi^{(0)}_k, \label{eq:sigma_k}
\end{eqnarray}
with 
\begin{eqnarray}
\sigma_{*k} = \frac{1+\tau-\tau\Gamma_k(1-\omega_{*i}/\omega)_k}{(1-\omega_{*e}/\omega)_k}, \label{eq:sigma_k_expression}
\end{eqnarray}
$\Gamma_k=I_0(b_k)\exp(-b_k)$ , $b_k = k_\perp^2 \rho_i^2$, $\rho_i^2 = (T_i/m_i)/\Omega_{ci}^2$,  and $I_0$ being the modified Bessel function. Noting  $\lvert k_{\perp}\rho_i\rvert \ll 1$   and   $\lvert \omega_{*i}/\omega\rvert\ll1$ for  most unstable TAEs,  one has $\sigma_{*k}\simeq 1$,  i.e., ideal MHD condition is satisfied at the lowest order.  At the next order, one has
\begin{eqnarray}
\delta H^{(1)}_{ki}\simeq \frac{e}{T_i}F_MJ_k\left(\delta\phi^{(1)}_k +\frac{\omega_{di}}{\omega}\delta\phi^{(0)}_k\right),  
\end{eqnarray}
with $\delta\phi^{(1)}_k$ being derived from quasi-neutrality condition, and contributing to SAW continuum upshift. Note that, here, we have dropped odd terms in $v_\parallel$ resulting in vanishing contributions to the dispersion response.  The hence obtained  particle response  can be substituted into linear gyrokinetic vorticity equation,  and yields, 
\begin{eqnarray}
\tau b_k \epsilon_{Ak} \delta\phi^{(0)}_k = 0, \label{eq:saw_dr_wkb}
\end{eqnarray}
with the SAW operator in the WKB limit given by
\begin{eqnarray}
\epsilon_{Ak}&\equiv& - \left(\frac{V^2_A}{b}\frac{k_{\parallel}b k_{\parallel}}{\omega^2}\right)_k\sigma_{*k} + \frac{1-\Gamma_k}{b_k}\left(1-\frac{\omega_{*i}}{\omega}\right)_k \nonumber\\
&&+ \left.\left\langle qJ_k\frac{\omega_d}{\omega} \delta H^{(1)}_{ki} \right\rangle \right/\left(\frac{n_0 e^2}{T_i} b_k\delta\phi^{(0)}_k\right).
\label{eq:SAW_operator}
\end{eqnarray}
The terms of $\epsilon_{Ak}$  correspond to field line bending, inertia and curvature coupling terms,  where ballooning-interchange terms are included. Resonant excitation by EPs can be straightforwardly accounted for by substituting the corresponding EP response into the curvature coupling term.   Note  that $b_k\equiv \rho^2_i \nabla^2_{\perp}$ and $k_{\parallel}$ should be strictly understood as operators, and are not commutative.  
The SAW instability wave equation and eigenmode dispersion relation in torus  can be derived by     transforming  equation (\ref{eq:saw_dr_wkb})  into ballooning space, and noting the two scale structure of SAW instabilities due to plasma nonuniformity.   Here, for simplicity of discussion, we focus  on modes in  the TAE frequency range  and,  thus,  the curvature coupling term that contributes to SAW continuum upshift and  BAE generation  is neglected from now on.   The  $\lvert\omega_{*i}/\omega\rvert$  correction is also systematically neglected except when explicitly stated and needed.   The perturbed scalar potential $\delta\phi_k$ can be decomposed as
\begin{eqnarray}
\delta\phi_k = A_k e^{-in\xi-i\omega t+im_0\theta} \sum_j e^{ij\theta}\Phi_j(nq-m),
\end{eqnarray}
with $A_k$ being the radial envelope, $m_0$ being the reference poloidal mode number, $m=m_0+j$, and $\Phi_j$ being the fine radial scale structure associated with $k_{\parallel}$. Defining $z=nq-m=-k_{\parallel} qR_0$,   $\eta$  being  the Fourier conjugate of $z$, and
\begin{eqnarray}
\Phi(z)=\int  \phi(\eta) e^{-i\eta z} d\eta,
\end{eqnarray}
the SAW eigenmode equation, equation  (\ref{eq:saw_dr_wkb}), can be reduced to the following simplified form for a $(\hat{s},\alpha)$ model equilibrium with shifted circular magnetic flux surfaces:
\begin{eqnarray}
\left[\frac{\partial^2}{\partial\eta^2} +\Omega^2_A\left(1+2\epsilon_0\cos\eta\right) - \frac{(\hat s - \alpha \cos \eta)^2}{\hat \kappa_\perp^4} + \frac{\alpha \cos \eta}{\hat \kappa_\perp^2}\right]\hat{\Phi}=0, \label{eq:SAW_eta}
\end{eqnarray}
with $\hat{\kappa}_{\perp}^2= - r^2\nabla_\perp^2/(n^2 q^2) = \left(\hat s\eta-\alpha \sin \eta\right)^2\left(1+2 \Delta^{\prime} \cos \eta\right) -2 \hat s\eta \Delta^{\prime} \sin \eta+1-2\left(r / R_0+\Delta^{\prime}\right) \cos \eta$, 
 $\hat{\Phi}\equiv \hat{\kappa}_{\perp}\phi(\eta)$,  $\hat{s}\equiv r q'/q$ being the magnetic shear and $\alpha = - R_0q^2 d\beta/dr$ the usual ballooning mode normalized pressure gradient,   $\Omega^2_A=\omega^2 q^2R^2_0/V^2_A$ ,  and $\epsilon_0=2(r/R_0+\Delta')$ with $\Delta'$ being Shafranov shift.  Equation (\ref{eq:SAW_eta}) has a clear two-scale character, and can be solved by asymptotic matching of two scale structures.  For inertial layer contribution with $\lvert\hat{s}\eta\rvert\gg1$ (often referred to as ``external region", denoted hereafter by the subscript E), equation (\ref{eq:SAW_eta}) reduces to
\begin{eqnarray}
\left[\frac{\partial^2}{\partial\eta^2} +\Omega^2_A\left(1+2\epsilon_0\cos\eta\right)  \right]\hat{\Phi}_E=0, \label{eq:SAW_Mathieu}
\end{eqnarray}
i.e., Mathieu's equation describing mode propagating in periodic systems, which can be solved noting its two scale character, 
\begin{eqnarray}
\hat{\Phi}_E =A(\eta)\cos(\eta/2) + B(\eta)\sin(\eta/2),\label{eq:phi_E}
\end{eqnarray} 
with $A(\eta)$ and $B(\eta)$ being slowly varying with respect to periodic variations reflecting typical $\lvert k_\parallel\rvert \simeq 1/2q R_0$ structures of TAE modes. One then has
\begin{eqnarray}
-B'(\eta)&=&\left(\Omega^2_A-1/4+\epsilon_0\Omega^2_A\right) A\equiv \Gamma_l A,\\
A'(\eta)&=& \left(\Omega^2_A-1/4-\epsilon_0\Omega^2_A\right) B\equiv  \Gamma_u B, 
\end{eqnarray}
with $\Gamma_{l}\equiv \Omega^2_A-1/4 + \epsilon_0\Omega^2_A$ and  $\Gamma_{u}\equiv \Omega^2_A-1/4 -  \epsilon_0\Omega^2_A$ determining the lower and upper accummulational points of toroidicity induced SAW continuum gap \cite{CZChengAP1985}, which then yields,
\begin{eqnarray}
\hat{\Phi}_E(\eta)=a\left(\sqrt{-\Gamma_u} \cos\frac{\eta}{2} \pm \sqrt{\Gamma_l}\sin\frac{\eta}{2}\right) e^{\mp \sqrt{-\Gamma_l\Gamma_u}\eta}.\label{eq:TAE_inertial_test}
\end{eqnarray}
The ``$\pm$" sign should be chosen in the way such that $e^{\mp \sqrt{-\Gamma_l\Gamma_u}\eta}$ decay as $\lvert\eta\rvert\rightarrow \infty$.  Noting equation (\ref{eq:phi_E}) and that $\eta$ is the Fourier conjugate of $z=-k_{\parallel} qR_0$, the $\cos(\eta/2)$- and $\sin(\eta/2)$-dependence of $\hat{\Phi}_E$ corresponds  to radial mode localization at $\lvert nq-m\rvert=1/2$;  i.e., the two neighbouring poloidal harmonics $m$ and $m\pm1$ couple  between two adjacent  mode rational surfaces, where $(nq - m) = - (nq - m\pm 1)$ and their respective dispersion relations are degenerate, forming the well-known ``rabbit-ear" like mode structure.  This feature of the radial TAE mode structure is important for the nonlinear mode coupling processes investigated in Sec. \ref{sec:TAE_nl}, due to the dominant contribution from the radially fast varying fluctuation structures in the inertial layer. The SAW continuum with corrections due to toroidicity,  can be obtained   from
\begin{eqnarray}
k_{\parallel} qR_0=\frac{1}{2}\pm\sqrt{\Gamma_l\Gamma_u},\label{eq:TAE_gap}
\end{eqnarray}
which then yields the toroidicity induced SAW continuum gap formation, inside which the discrete TAE can be excited with minimized continuum damping.  A sketched continuum is shown in Fig.  \ref{fig:TAE_gap}.  The corresponding discrete Alfv\'en eigenmode, i.e., TAE,  can then be excited by, e.g., EPs, inside this toroidicity induced continuum gap, with minimum  requirement on EP drive due to the minimized continuum damping \cite{LChenVarenna1988,GFuPoFB1989}.  The TAE excitation mechanism, however, is beyond the scope of the present review, focusing on the nonlinear evolution of TAE with prescribed amplitude due to nonlinear wave-wave coupling, and will not be addressed here. 

\section{TAE saturation via nonlinear wave-wave coupling}\label{sec:TAE_nl}

Nonlinear mode coupling describes the TAE distortion  due to    interaction with other oscillations, and is expected to play crucial role in TAE nonlinear saturation in future reactors, where system size will be much larger than characteristic orbit size of EPs, and,  thus,  SAW instabilities with a broad spectrum in toroidal mode numbers  will  be simultaneously excited by EPs.  To illustrate the richness of nonlinear mode couplings of TAE and the  strength  of gyrokinetic theory in the investigation of underlying physics, three examples are presented, i.e., the nonlinear excitation of  $n=0$ zonal field structure (ZFS)  by TAE \cite{LChenPRL2012}, which corresponds to single-n TAE nonlinear envelope modification  via modulational instability;  nonlinear spectral evolution of TAE via ion induced scattering \cite{TSHahmPRL1995,ZQiuNF2019a}, which is expected to play crucial role in determining the broad toroidal mode number  TAE saturated spectrum and ensuing EP transport; and cross-scale scattering and damping  of meso-scale TAE by micro-scale  DW \cite{LChenNF2022},  suggested  by recent experiments  as well as simulations showing  improved thermal plasma confinement in the presence  of significant amount of  EPs \cite{JCitrinPRL2013,ADiSienaNF2019,SMazziNP2022}.  All these three   channels of wave-wave couplings  are shown to significantly influence the TAE nonlinear dynamics in  different  ways, and their relative importance  and implications on TAE saturation in burning plasma parameter regimes  are discussed.   For the sake of simplicity and for consistency with original literature, different notations that are needed  are defined  only in the corresponding subsection.

\subsection{ZFS generation by TAE}
\label{sec:ZFS_excitation}

Zonal field structures correspond to toroidally and near poloidally  symmetric perturbations with $n=0$, and are, thus, linearly stable since they cannot tap the expansion free energy associated with plasma profile nonuniformities.  ZFS can be nonlinearly excited by  DW  turbulence  including drift Alfv\'en waves (DAWs), and in this process, self-consistently scatter DW/DAW into the linearly stable short radial wavelength domain, leading to   turbulence regulation and confinement improvement. ZFS excitation was extensively studied in the  DWs dynamics \cite{ZLinScience1998,LChenPoP2000,FZoncaPoP2004,PDiamondPPCF2005},  observed in simulations with TAEs \cite{DSpongPoP1994,YTodoNF2010},   while theoretical implications of ZFS to the nonlinear physics of TAE  were discussed in \cite{LChenPRL2012}.  The nonlinear excitation process can be described by the four-wave modulational instability, where upper/lower TAE sidebands due to ZFS modulation are generated, and the nonlinear dispersion relation for ZFS generation can be obtained by the coupled ZFS and TAE sidebands equations. It is noteworthy that, both electrostatic zonal flow (ZF) and electromagnetic zonal magnetic field (zonal current, ZC) should be accounted for on the same footing for the proper understanding of the ZFS generation process \cite{FZoncaEPL2015,LChenPRL2012}. 

\begin{figure}
\includegraphics[scale=0.40]{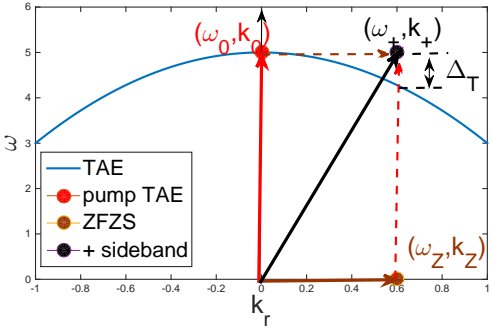} 
\caption{Frequency and wavenumber matching condition for ZFS generation by TAE. Here, the horizontal axis is the radial envelope wavenumber $k_r$, and vertical axis is the frequency. The solid curve is the TAE dispersion relation, and $\Delta_T$ is the frequency mismatch.}\label{fig:MI_coupling}
\end{figure}

For the clarity of presentation, we  focus on  the modulational instability of TAE originally investigated in Ref. \cite{LChenPRL2012}. The further extensions, including the enhanced nonlinear coupling due to existence of ``fine-radial-scale" structure ZFS \cite{ZQiuNF2017}, and effects of resonant EPs in rendering the ZFS generation process into a forced driven process \cite{ZQiuPoP2016}  will be  only briefly discussed at the end of this section  to give the readers a complete picture of the state-of-art research. Considering that TAE constitutes the pump wave $\Omega_0(\omega_0,\mathbf{k}_0)$ and its upper and lower sidebands $\Omega_\pm(\omega_{\pm},\mathbf{k}_{\pm})$ due to the radial modulation of ZFS $\Omega_Z(\omega_Z,\mathbf{k}_Z)$, and assuming  $\Omega_{\pm}=\Omega_Z\pm\Omega_0$ as the wave vector/frequency matching conditions, the perturbations can be decomposed as
\begin{eqnarray}
\delta\phi_0 &=& A_0 e^{i(n\phi-m_0\theta-\omega_0 t)} \sum_j e^{-ij\theta}\Phi_0(x-j),\nonumber\\
\delta\phi_{\pm}&=& A_{\pm} e^{\pm(n\phi-m_0\theta-\omega_0t)} e^{i(\int k_Z dr -\omega_Z t)}    \sum_j e^{\mp ij\theta}\left\{\Phi_0(x-j)\atop\Phi^*_0(x-j)\right\},\nonumber\\
\delta\phi_Z&=& A_Z e^{i(\int k_Z dr-\omega_Z t)}. \nonumber
\end{eqnarray}
 The frequency and wavenumber matching conditions are already assumed, as illustrated in Fig. \ref{fig:MI_coupling}.   We note that, the expression of $\delta\phi_{\pm}$ indicates that the  parallel mode structure ($\Phi_0$) is not altered by the radial envelope modulation process, which occurs on a longer time scale than the formation of the parallel mode structure itself. 

We start from ZFS generation. The first equation for zonal flow generation can be derived from nonlinear vorticity equation.  Noting that ZFS have $k_{\parallel Z}=0$,  one obtains
\begin{eqnarray}
&&\frac{n_0e^2}{T_i}\left\langle (1-J^2_Z)\frac{F_M}{n_0}\right\rangle\delta\phi_Z - \sum_{s=e,i} \left\langle \overline{\frac{q}{\omega}J_Z \omega_d\delta H^{(1)}_Z}\right\rangle\nonumber\\
&=&-\frac{i}{\omega_Z}\sum_{\mathbf{k}'+\mathbf{k}''=\mathbf{k}_Z} \Lambda^{k'}_{k''} \left[\left\langle e(J_ZJ_{k'}-J_{k''}) \delta L_{k'}\delta H_{k''}\right\rangle +\frac{c^2}{4\pi} k''^2_{\perp}  \frac{\partial_l\delta\psi_{k'}\partial_l\delta\psi_{k''}}{\omega_{k'}\omega_{k''}} \right].\nonumber
\end{eqnarray}
Substituting ion responses of $\Omega_0$ and $\Omega_{\pm}$ into RS,  noting  $k_{\perp}\rho_i\lesssim O(1)$, and averaging over fast varying radial scale, one obtains
\begin{eqnarray}
i\omega_Z\hat{\chi}_{iZ} \delta\phi_Z=-\frac{c}{B_0}k_{\theta0}k_Z \left(1-\frac{\omega^2_A}{4\omega^2_0}\right)\left(A_0A_--A_{0^*}A_+\right).\label{eq:ZF_TAE}
\end{eqnarray}
Here, $\hat{\chi}_{iZ}\equiv \chi_{iZ}/(k^2_Z\rho^2_i)$, with $\chi_{iZ}\simeq 1.6 k^2_Z\rho^2_iq^2/\sqrt{\epsilon}$ corresponds to the neoclassical inertial enhancement \cite{MRosenbluthPRL1998}, $\omega_A=V_A/(qR_0)$ and   $1-\omega^2_A/(4\omega^2_0)\sim O(\epsilon)$ corresponds to the RS and MX non-cancellation  due to toroidicity, and finite coupling comes from radial envelope modulation ($\propto k^2_Z\rho^2_i$) by ZFS.

The zonal magnetic field equation  can be derived from electron parallel force balance equation in stead of  the quasi-neutrality condition,
\begin{eqnarray}
\delta E_{\parallel}+\mathbf{b}\cdot\delta \mathbf{u}_{\perp}\times\delta \mathbf{B}_{\perp}/c=0.\label{eq:force_balance}
\end{eqnarray}
Noting that $\delta E_{\parallel}\equiv -\partial_l\delta\phi-c\partial_t\delta A_{\parallel}$, $\delta \mathbf{u}_{\perp}\simeq c\mathbf{b}\times \nabla_{\perp}\delta\phi/B$, $\delta\mathbf{B}_{\perp}=\nabla\times \delta A_{\parallel}\mathbf{b}$,  one  also obtains,  
\begin{eqnarray}
\delta\psi_Z=-\frac{i}{\omega_0}\frac{c}{B_0}k_Zk_{\theta0}\left(A_0A_-+A_{0^*}A_+\right),\label{eq:ZC_TAE}
\end{eqnarray} 
where we have introduced the parallel induction potential $\delta \psi_Z = \omega_0 \delta A_{\parallel Z}/(c k_{\parallel 0})$ for ZFS by analogy to the definition of $\delta \psi_k$ for TAE. In deriving  equation (\ref{eq:ZC_TAE}), we also noted $\omega_{\pm}=\omega_Z\pm\omega_0$ as well as ideal MHD condition for TAEs. 

The TAE sidebands equations can be derived from nonlinear vorticity equation. We will start with the upper sideband, while the derivation of the  governing equations for the lower sideband is similar.   Neglecting the curvature coupling term due to the $\lvert\omega\rvert\gg\omega_G$ ordering for TAEs,  substituting the linear ion responses to $\Omega_0$ and $\Omega_Z$ into equation (\ref{eq:vorticity}), and noting $k_{\perp}\rho_i\lesssim O(1)$, we have
\begin{eqnarray}
k^2_{\perp+}\left[-k^2_{\parallel0}\delta\psi_+ + \frac{\omega^2_+}{V^2_A}\delta\phi_+\right] =-i\frac{c}{B_0} k_Zk_{\theta0}  \left(k^2_Z-k^2_{\perp0}\right)\frac{\omega_0}{V^2_A} \delta\phi_0\left(\delta\phi_Z-\delta\psi_Z\right).\label{eq:TAE_ZF_us_v1}
\end{eqnarray}

The other equation for $\Omega_+$ can be derived from the electron  parallel force balance equation, equation (\ref{eq:force_balance}), noting that $k_{\parallel0}=k_{\parallel+}$ and $\delta\phi_0\simeq\delta\psi_0$ for the pump TAE, and we obtain:  
\begin{eqnarray}
\delta\phi_+-\delta\psi_+=i\frac{c}{B_0}k_Zk_{\theta0} \frac{1}{\omega_0}\delta\phi_0\left(\delta\psi_Z-\delta\phi_Z\right). \label{eq:TAE_ZF_us_qn}
\end{eqnarray}

Substituting equation (\ref{eq:TAE_ZF_us_qn}) into (\ref{eq:TAE_ZF_us_v1}), one then have 
\begin{eqnarray}
b_+ \epsilon_{A+} \delta\phi_+ = 2 \frac{i}{\omega_0} \frac{c}{B_0}  k_{\theta0}k_Z b_0 \delta\phi_0\left(\delta\phi_Z-\delta\psi_Z\right),\label{eq:TAE_ZF_us_wkb}
\end{eqnarray}
with $\epsilon_{A+}$ being the $\Omega_+$ dispersion relation in the WKB limit.  The $\Omega_-$ equation can be derived similarly. 
Multiplying  both sides of equation (\ref{eq:TAE_ZF_us_wkb}) by $\Phi_0$ and averaging over the  fast  radial   scale,   one has
\begin{eqnarray}
b_{\pm}\hat{\epsilon}_{A\pm}A_{\pm}=2  \frac{i}{\omega_0} \frac{c}{B_0}  k_{\theta0}k_Z b_0 \left(A_0\atop A^*_0\right)\left(\delta\phi_Z-\delta\psi_Z\right),\label{eq:TAE_ZF_us_eigen}
\end{eqnarray}
with 
\begin{eqnarray}
\hat{\epsilon}_{A\pm}&=& \left(\omega^4_A \Lambda_T(\omega) D(\omega,k_Z)/\epsilon_0\right)_{\omega=\omega_{\pm}},\\
D(\omega,k_Z)&=& \Lambda_T(\omega)-\delta\hat{W}(\omega,k_Z),
\end{eqnarray}
$\Lambda_T\equiv \sqrt{-\Gamma_l\Gamma_u}$ as given by equation (\ref{eq:TAE_gap}), and $\delta\hat{W}(\omega,k_Z)$ being the normalized potential energy.

The modulational dispersion relation for ZFS generation by TAE  can then be derived from equations (\ref{eq:ZF_TAE}), (\ref{eq:ZC_TAE}), and (\ref{eq:TAE_ZF_us_eigen}), and one obtains
\begin{eqnarray}
&&2\left(\frac{c}{B_0}k_Zk_{\theta0}\lvert A_0\rvert\right)^2\frac{b_0}{b_Z}\left[\frac{1-\omega^2_A/(4\omega^2_0)} {\hat{\chi}_{iZ} (\omega_Z/\omega_0)}  
\left(\frac{1}{\hat{\epsilon}_{A+}}- \frac{1}{\hat{\epsilon}_{A-}}\right)  +   \left(\frac{1}{\hat{\epsilon}_{A+}}+ \frac{1}{\hat{\epsilon}_{A-}}\right) \right]\nonumber\\
&&=-1, \label{eq:modu_dr}
\end{eqnarray}
 which can be solved by expanding $D(\omega_{\pm},k_Z)$ as
\begin{eqnarray}
D(\omega_{\pm},k_Z)=\pm \frac{\partial D}{\partial\omega_0}\left(i\gamma_Z\mp\Delta_T\right),
\end{eqnarray}
with $\gamma_Z=-i\omega_Z$ and $\Delta_T\equiv \omega_T(k_Z)-\omega_0$ being the frequency mismatch as shown in Fig. \ref{fig:MI_coupling}, and one obtains
\begin{eqnarray}
\gamma^2_Z&=&\left(\frac{c}{B_0}k_Zk_{\theta0}\lvert A_0\rvert\right)^2\frac{b_0}{b_Z}\frac{\epsilon_0}{\Lambda_T}\frac{4\omega_0/\omega^2_A}{\partial D_0/\partial\omega_0}\left[\frac{\Delta_T}{\omega_0}\frac{\omega^2_0}{\omega^2_A} + \frac{1}{\hat{\chi}_{iZ}}\left(\frac{\omega^2_0}{\omega^2_A}-\frac{1}{4}\right)\right]-\Delta^2_T,\nonumber\\
&&
\end{eqnarray}
with the first term in the square brackets ($\propto \Delta_T/\omega_0$) corresponding to the contribution from ZC, while the other term accounts for ZF contribution. It is readily seen that   ZF contribution can be of higher order due to the neoclassical shielding ($1/\hat{\chi}_{iZ}\ll1$) and RS-MX near cancellation by $\omega^2_0/\omega^2_A-1/4\sim O(\epsilon)$. Thus,  for $\Delta_T/\omega_0>0$,   excitation via the ZC channel  can be favoured  due  to its much lower threshold condition on pump TAE amplitude $A_0$. On the other hand, for $\Delta_T/\omega_0<0$, ZF excitation is still possible, however, on quite stringent conditions;  i.e., $\omega^2_0/\omega^2_A>1/4$,  which corresponds to the pump TAE located in  the upper half of the toroidicity induced continuum gap \cite{ZQiuEPL2013}   and the pump TAE amplitude being large enough to overcome the threshold due to frequency mismatch.  It thus suggests that ZFS may be dominated by ZC for $\Delta_T/\omega_0 > 0$  due to the trapped-ion enhanced polarizability;  thus, a kinetic treatment is necessary. On the other hand, if MHD model without trapped particle effects is adopted, the  obtained ZFS excitation  condition and corresponding ZFS level  will  be qualitatively in-correct.  This discussion illustrates the richness of the phenomenology underlying the nonlinear route to TAE saturation via wave-wave coupling and ZFS generation. Meanwhile, it also clarifies that the result of numerical simulations may depend on the adopted physics model. Furthermore, it is also worth noting that the ``preferential channel via ZC excitation" is connected with the properties of  the TAE of interest here, for which RS and MX nearly cancel each other.  This argument cannot be straightforwardly generalized to other SAW instabilities;  e.g.,  BAE with $\lvert k_{\parallel}V_A/\omega\rvert\ll1$ will predominantly excite ZF \cite{ZQiuNF2016,HZhangPST2013}; while for reversed shear Alfv\'en eigenmode (RSAE) with frequency between TAE and BAE frequency range, depending on the specific $\lvert nq_{min}-m\rvert$ value, both ZF and/or ZC  excitation can be preferred \cite{SWeiJPP2021}.

For ZC excitation with $\Delta_T/\omega_0>0$ and typical parameters of most unstable TAE with $k_{\perp}\rho_E\sim O(1)$, the threshold condition can be estimated as 
\begin{eqnarray}
\left\lvert \frac{\delta B_{r0}}{B_0}\right\rvert^2\sim O(10^{-8}\sim 10^{-7}),
\end{eqnarray}
which is consistent with the observed magnetic perturbations in present day tokamak experiments \cite{WHeidbrinkPRL2007}, suggesting the ZFS excitation can be important for TAE saturation. As the drive by pump TAE is significantly higher than the threshold,   the ZFS growth rate  is linearly proportional to pump TAE amplitude, typical of spontaneous excitation processes by modulational instability. This feature identifies the parameter space region, where spontaneous excitation is dominant, and clearly distinguishable from,  e.g., the forced driven process with the ZFS growth rate determined by the instaneous TAE growth rate, as  discussed below \cite{ABiancalaniPPCF2021,ZQiuPoP2016}.

In the present analysis, only  thermal plasma contribution to inertial layer is considered; consistent with EP contribution being negligible in the   perpendicular scattering  process   due to the $k_{\perp}\rho_E\gg 1$ ordering. The EP response, however, can play an important role in  the ideal region, as addressed in Ref. \cite{ZQiuPoP2016}, where it was shown that, as the pump TAE is exponentially growing due to resonant EP contribution,  nonlinear EP response to ZFS   contributes to  the curvature-pressure  term in the vorticity equation,   dominanting over the RS and MX in the uniform plasma limit. This EP enhanced coupling occurs in the exponentially growing stage of the pump TAE, with ZF excitation dominating over ZC contribution. In that case,  the ZF excitation process corresponds  to a ``forced driven" process, with the ZF growth rate being twice of the instaneous TAE growth rate, as frequently observed in numerical simulations \cite{YTodoNF2010,ABiancalaniPPCF2021,SMazziNP2022}.

\begin{figure}
\includegraphics[scale=0.35]{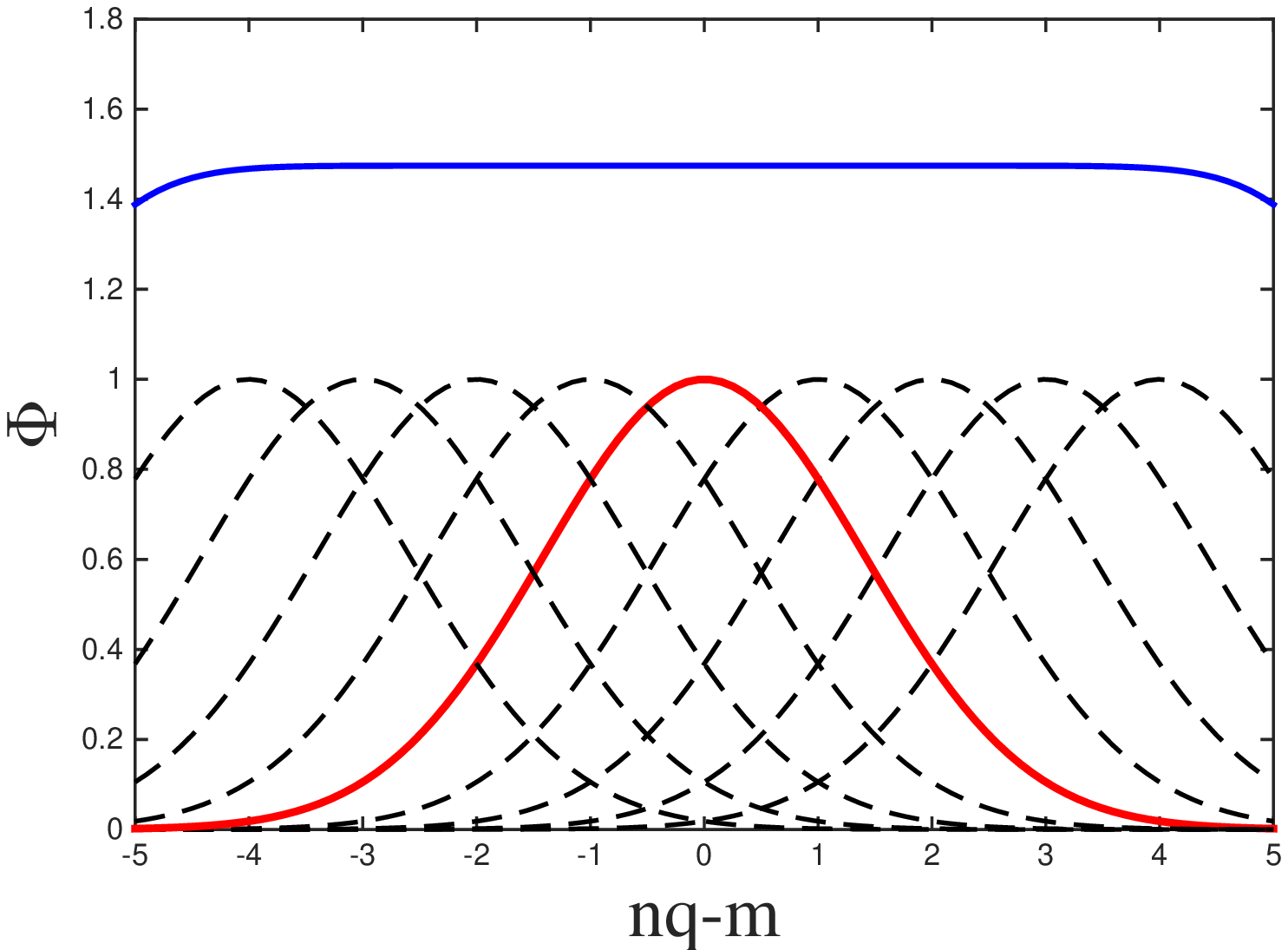} 
\includegraphics[scale=0.30]{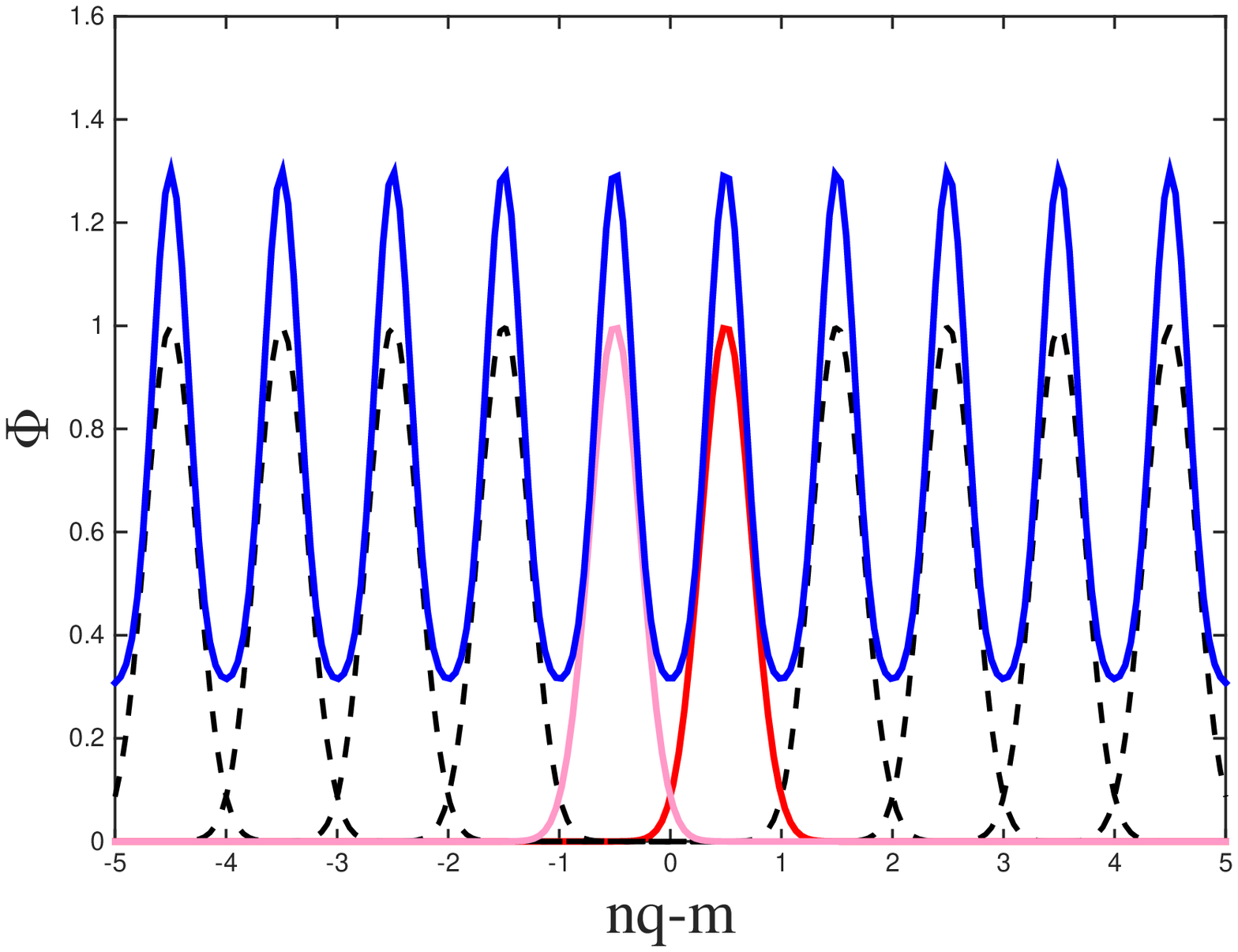} 
\caption{Cartoon for ZFS excitation by strongly ballooning DWs (left panel) v.s. weakly ballooning SAW instabilities (right panel). Here, the dashed curves correspond to the parallel mode structure $\Phi_0(nq-m)$ for DWs (left panel) and SAW instabilities (right panel), respectively;  while the solid blue curves in both panels correspond to $\sum_m\lvert \Phi_0\rvert^2$. Thus, for DWs with $\sum_m\lvert\Phi_0\rvert$  being almost independent of $r$ \cite{FZoncaPoP2004}, radial envelope modulation leads to meso-scale ZF excitation; while for SAW instabilities,  fine-scale structure ZFS is excited. }\label{fig:ballooning}
\end{figure}

Another important finding on ZFS excitation  by SAW instabilities is  due to their  weak/moderate ballooning features,  corresponding to a smaller or comparable radial width of the parallel mode structure with respect to  the   distance between mode rational surfaces. As a result,  the ZFS excited by TAE has   a   fine-scale radial    structure \cite{ZQiuNF2016,HZhangPST2013},  in addition to the meso-scale
radial envelope corrugation, different from the well-known ``meso"-scale ZF excitation in the typically moderately/strongly ballooning DWs,  as shown in Fig. \ref{fig:ballooning}. This fine-scale radial  structure may significantly enhance the ZFS generation rate and its impact on regulating SAW instabilities via the perpendicular scattering.  For a comprehensive review of gyrokinetic  theory of ZFS generation by TAE, interested readers may refer to Ref. \cite{ZQiuNF2017},  where different physics are clarified;  e.g., forced driven v.s. spontaneous excitation, meso-scale corrugation v.s. fine-scale structure.

\subsection{TAE saturation due to ion induced scattering}

Nonlinear ion induced scattering is another potentially  important channel for SAW instability nonlinear saturation, corresponding  to   parametric decay into another SAW and a heavily ion Landau damped ion quasi-mode \cite{RSagdeevbook1969}. The role of this process in TAE saturation  was originally explored in Ref. \cite{TSHahmPRL1995}.  It is of particular interest since TAEs lie between two neighbouring mode rational surfaces and are characterized by finite parallel wavenumber $\lvert k_{\parallel}\rvert\simeq 1/(2qR_0)$, as discussed in Sec. \ref{sec:saw_torus}. Thus, as two counter-propagating  TAEs couple, a low frequency mode with finite parallel wavenumber can be generated, i.e., an ion quasi-mode, which can be heavily ion Landau damped, leading to significant consequence on TAE nonlinear dynamics. Compared to ZFS generation investigated in the previous section as a self-interaction process of a single-n 
TAE, ion induced scattering process is expected to be of particular importance in reactor scale machines with system size being much larger than the characteristic  orbit width of fusion alpha particles,  where   TAEs  with multiple toroidal mode numbers and comparable linear growth rates could 
coexist  \cite{LChenRMP2016,TWangPoP2018,ZRenNF2020}. Thus, the ion induced scattering process  can determine the saturated spectrum of TAEs and the consequent alpha particle transport rate.  Meanwhile, the   Landau damping of the nonlinearly generated ion quasi-mode will indirectly transfer the fusion alpha particle power to  heat  deuterium and tritium ions, providing a potential effective alpha-channeling mechanism \cite{TSHahmPST2015,NFischNF1994,NFischPRL1992,ZQiuPRL2018,ZQiuNF2019b,SWeiNF2022}.

\begin{figure}
\includegraphics[scale=0.35]{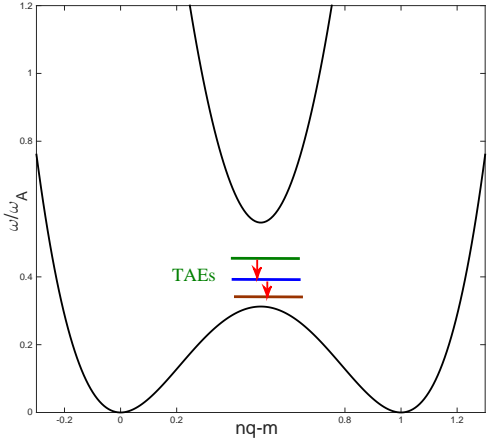} 
\caption{Cartoon of TAE parametric decay in the low-$\beta$ limit.}\label{fig:TAE_ISW_TAE}
\end{figure}

The TAE saturation via ion induced scattering  was originally investigated in Ref. \cite{TSHahmPRL1995} using drift kinetic theory, which was generalized  to fusion relevant  short wavelength regime  with $k^2_{\perp}\rho^2_i\gg \lvert\omega/\Omega_{ci}\rvert$ in Ref. \cite{ZQiuNF2019a}. Correspondingly,  the dominant  nonlinear scattering mechanism is qualitatively replaced by the perpendicular scattering \cite{LChenEPL2011}, and the saturation level is consequently reduced by one order of magnitude.  
However, the conceptual workflow  of Ref. \citenum{ZQiuNF2019a} is similar to that of Ref. \cite{TSHahmPRL1995}. In a single scattering process, a pump TAE decays into a counter-propagating  sideband TAE and an ion quasi-mode,  and the parametric decay process can  occur spontaneously as the sideband TAE frequency is lower than that of the pump wave, as shown in Fig. \ref{fig:TAE_ISW_TAE}. 
This process may lead to TAE saturation as the sideband TAE is  damped due to the enhanced coupling to lower accumulational point of  SAW continuum.
As there are many TAEs co-existing, each TAE may simultaneously  interact with many TAEs;  in some processes it may act as the pump wave, while in some other processes it acts as the decay wave. To  analyze this spectral cascading process, the interaction of a  representative  ``test" TAE with a ``background" TAE is studied; and the equation for the test TAE nonlinear evolution due to interaction with the background TAE   and the ion quasi-mode is derived.  In the case of  multiple background TAEs simultaneously interacting with the test TAE,  one then obtains the equation describing TAE spectral evolution.  

Thus, with the linear instability spectrum determined by the equilibrium profiles, the nonlinear process  gives the nonlinear saturation spectrum, which eventually determines the electromagnetic fluctuation induced alpha particle transport, as sketched in Fig. \ref{fig:TAE_cascading}.

\subsubsection{Parametric decay instability}

Starting from the nonlinear interaction of the test TAE $\Omega_0(\omega_0,\mathbf{k}_0)$  with the counter-propagating  background TAE $\Omega_1(\omega_1,\mathbf{k}_1)$, during which the ion sound mode (ISM) $\Omega_s(\omega_s,\mathbf{k}_s)$ fluctuation is generated, our analysis involves the coupled equations of ISM generation and background TAE evolution.  Considering the  $\lvert k_{\parallel s}v_e\rvert\gg\lvert \omega_s\rvert, \lvert\omega_{d s}\rvert$ ordering, and assuming electrostatic ISM,  the linear thermal plasma response to ISM can be derived as  
\begin{eqnarray}
\delta H^{(1)}_{si}&=&\frac{e}{T_i} F_{M} \frac{\omega_s}{\omega_s-k_{\parallel s}v_{\parallel}} J_s\delta\phi_s,\\
\delta H^{(1)}_{se}&=&0.
\end{eqnarray}

Adopting  the linear electron response to TAEs derived in equation (\ref{eq:linear_e_TAE}),  the nonlinear gyrokinetic equation for electron response to ISW becomes
\begin{eqnarray}
v_{\parallel}\partial_l\delta H^{(2)}_{s e} &=& - \sum_{\mathbf{k}'+\mathbf{k}''=\mathbf{k}} \Lambda^{k'}_{k''}  \delta L_{k'}\delta H_{k'' e}\nonumber\\
&\simeq& -\Lambda^{k^*_1}_{k_0} \frac{e}{T_e} F_M v_{\parallel} \left(\frac{k_{\parallel 1^*}}{\omega_{1^*}}- \frac{k_{\parallel0}}{\omega_0}\right)\delta \phi_0\delta\psi_{1^*}, \label{eq:nl_e_ISW}
\end{eqnarray}
with $ \Lambda^{k^*_1}_{k_0}\equiv (c/B_0) \hat{\mathbf{b}}\cdot\mathbf{k_0}\times\mathbf{k_{1^*}}$.   Noting that $\omega_{1^*}\simeq -\omega_0$, $k_{\parallel 1^*}\simeq k_{\parallel 0}$ and consequently that $k_{\parallel s}\simeq 2k_{\parallel 0}$, one has
\begin{eqnarray}
\delta H^{(2)}_{se}\simeq -i\frac{\Lambda^{k^*_1}_{k_0}}{\omega_0}\frac{e}{T_e} F_M\delta \phi_0\delta\psi_{1^*}.
\end{eqnarray}
Nonlinear ion response to $\Omega_s$ can be derived noting the $\omega_s\sim k_{\parallel s}v_{\parallel}\gg\omega_{d s}$ ordering, and one has
\begin{eqnarray}
\delta H^{(2)}_{si}\simeq -i\frac{\Lambda^{k^*_1}_{k_0}}{\omega_0}\frac{e}{T_i} F_M\frac{k_{\parallel s}v_{\parallel}}{\omega_s-k_{\parallel s}v_{\parallel}} J_0J_1\delta\phi_0\delta\phi_{1^*}. \label{eq:nl_i_ISW}
\end{eqnarray}
It is noteworthy that  $\omega_s\sim k_{\parallel s}v_{\parallel}$, which  is crucial for the resonant wave-particle interactions that determines the scattering process.  Substituting equations (\ref{eq:nl_e_ISW}) and (\ref{eq:nl_i_ISW}) into quasi-neutrality condition, one obtains the nonlinear equation for $\Omega_s$ generation:
\begin{eqnarray}
\epsilon_s\delta\phi_s=i\frac{\Lambda^{k^*_1}_{k_0}}{\omega_0}\beta_1 \delta\phi_0\delta\phi_{1^*},\label{eq:nl_isw}
\end{eqnarray}
with $\epsilon_s\equiv 1+\tau+\tau\Gamma_s\xi_s Z(\xi_s)$ being the ISM linear dispersion relation, $\xi_s\equiv \omega_s/(k_{\parallel s} v_{it})$,   $Z(\xi_s)$ being the well-known plasma dispersion function defined as 
\[Z(\xi_s) \equiv\frac{1}{\sqrt{\pi}} \int^{\infty}_{-\infty} \frac{e^{-y^2}}{(y-\xi_s)} dy,  \]
  the nonlinear coupling coefficient $\beta_1=1+\tau F_1 (1+\xi_sZ(\xi_s))$ and $F_1\equiv \langle J_0J_1J_sF_{M}/n_0\rangle$. 

The nonlinear particle response to the test TAE,  can be derived as 
\begin{eqnarray}
\delta H^{(2)}_{0e}&=& -\frac{(\Lambda^{k^*_1}_{k_0})^2}{\omega^2_0} \frac{e}{T_e} F_M \lvert\delta\phi_1\rvert^2\delta\phi_0,\\
\delta H^{(2)}_{0i}&=& i\frac{\Lambda^{k^*_1}_{k_0}}{\omega_0} \frac{e}{T_i} F_M \frac{k_{\parallel s}v_{\parallel}}{\omega_s-k_{\parallel s}v_{\parallel}} \left[J_1J_s\delta\phi_s\delta\phi_1 -i\frac{\Lambda^{k^*_1}_{k_0}}{\omega_0} J^2_1 J_0\lvert\delta\phi_1\rvert^2\delta\phi_0 \right].
\end{eqnarray}
In deriving $\delta H^{(2)}_{0e}$ and $\delta H^{(2)}_{0i}$,  the nonlinear particle responses to $\Omega_s$ are also included due to the fact that it may be heavily ion Landau damped.  One then obtains,
\begin{eqnarray}
\delta\psi_0=\left(1+\sigma^{(2)}_0\right)\delta\phi_0 + D_0\delta\phi_1\delta\phi_s,\label{eq:nl_qn_test}
\end{eqnarray}
with $\sigma^{(2)}_0\equiv (\Lambda^{k^*_1}_{k_0})^2 \left[-1+\tau F_2 (1+\xi_s Z(\xi_s)\right] \lvert\delta\phi_1\rvert^2/\omega^2_0$, $D_0=i\tau \Lambda^{k^*_1}_{k_0} F_1 [1+\xi_sZ(\xi_s)]/\omega_0$, and $F_2=\langle J^2_0J^2_1 F_M/n_0\rangle$.  

The other equation of $\Omega_0$ can be derived from nonlinear vorticity equation as
\begin{eqnarray}
b_0\left[ \frac{1-\Gamma_0+\alpha^{(2)}_0/\omega^2_0}{b_0} \delta\phi_0 -\frac{k^2_{\parallel0}V^2_A}{\omega^2_0} \delta\psi_0  \right] = D_2 \delta\phi_1\delta\phi_s,\label{eq:nl_vorticity_test}
\end{eqnarray}
with 
\begin{eqnarray}
\alpha^{(2)}_0&=&(\Lambda^{k^*_1}_{k_0})^2 (F_2-F_1) (1+\xi_s Z(\xi_s))\lvert\delta\phi_1\rvert^2,\nonumber\\
D_2&=& -i \Lambda^{k^*_1}_{k_0}[F_1(1+\xi_sZ(\xi_s)) -\Gamma_s\xi_sZ(\xi_s)-\Gamma_1]/\omega_0.\nonumber
\end{eqnarray}

From equations (\ref{eq:nl_qn_test}),  (\ref{eq:nl_vorticity_test}) and (\ref{eq:nl_isw}), one obtains the following nonlinear eigenmode equation of the test TAE $\Omega_0$ due to interaction  with the background TAE $\Omega_1$
\begin{eqnarray}
b_0\left(\epsilon_{A0}+\epsilon^{(2)}_0\right)\delta\phi_0 =- \frac{(\Lambda^{k^*_1}_{k_0})^2\beta_1\beta_2}{\tau\epsilon_s} \lvert\delta\phi_1\rvert^2\delta\phi_0,\label{eq:nl_test_wkb}
\end{eqnarray}
with $\beta_2\equiv \beta_1-\epsilon_s$.  Multiplying both sides of equation (\ref{eq:nl_test_wkb}) with $\Phi^*_0$, and averaging over the fast radial scale of $1/(n_sq')\ll \delta\ll 1/(n_0q')$, one then obtains
\begin{eqnarray}
\left(\hat{\epsilon}_{A0}-\Delta_0\lvert A_1\rvert^2-\chi_0\epsilon_s\lvert A_1\rvert^2\right) A_0=-(\hat{C}_0/\epsilon_s) \lvert A_1\rvert^2A_0, \label{eq:nl_test_eigen}
\end{eqnarray}
with $\hat{\epsilon}_{A0}$ being the $\Omega_0$ linear eigenmode dispersion relation obtained from $\hat{\epsilon}_{A0}\equiv \int \lvert\Phi_0\rvert^2\epsilon_{A0} dr/\int \lvert\Phi_0\rvert^2 dr$,  $\Delta_0$, $\chi_0$ and $\hat{C}_0$ corresponding, respectively,  to nonlinear frequency shift, ion Compton scattering and shielded-ion scattering.  Their specific  expressions  can be found in Ref. \cite{ZQiuNF2019a}.  Equation (\ref{eq:nl_test_eigen}) can be understood as the parametric   dispersion relation for $\delta\phi_1$ decaying into $\delta\phi_0$ and $\delta\phi_s$   and the condition for the nonlinear process to occur  can be determined for different parameter regimes that crucially enter through the properties of $\delta\phi_s$.

For typical tokamak parameters with $\tau\sim O(1)$, $\Omega_s$ is  heavily Landau damped with $\lvert\epsilon_{s,I}\rvert$ comparable to $\lvert\epsilon_{s,R}\rvert$, with subscripts ``$R$" and ``$I$" denoting real and imaginary parts. One then has, from the imaginary part of equation (\ref{eq:nl_test_eigen}),
\begin{eqnarray}
\gamma+\gamma_0=\frac{\lvert A_1\rvert^2}{\partial_{\omega_0}\epsilon_{0,R}} \left(\frac{\hat{C}_0}{\lvert\epsilon_{s}\rvert^2} +\chi_0\right) \epsilon_{s,I}.
\end{eqnarray}
with $\hat{C}_0$ and $\chi_0$ corresponding, respectively, to the shielded-ion and nonlinear ion Compton scatterings. Since both  $\hat{C}_0$ and $\chi_0$ are positive definite,  and that $\epsilon_{s,I}=\sqrt{\pi}\tau \Gamma_s\xi_s\exp(-\xi^2_s)$ with $\xi_s\equiv (\omega_0-\omega_1)/\lvert k_{\parallel s} v_{it}\rvert$, one then has, $\gamma>0$ corresponds to $\omega_1>\omega_0$, i.e., the parametric decay spontaneously occur as the pump TAE frequency is higher than the sideband TAE. Thus, the above discussed parametric decay process will lead to power transfer from higher frequency   part of the spectrum to the lower frequency part \cite{RSagdeevbook1969,TSHahmPRL1995}, as shown in Fig. \ref{fig:TAE_cascading}.  The  sideband TAE, with lower frequency,  can   be saturated due to enhanced   damping  due to coupling  to  the lower part of the SAW continuum.

\subsubsection{Spectral evolution}

\begin{figure}
\includegraphics[scale=0.40]{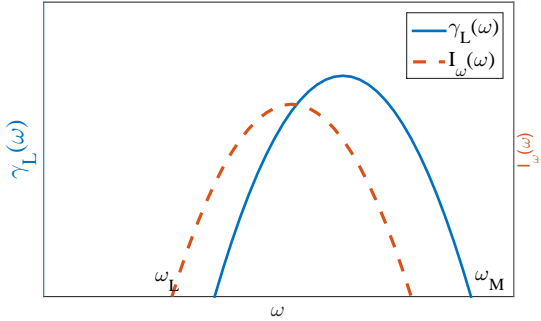} 
\caption{Cartoon of TAE spectral cascading due to ion induced scattering. The horizontal axis is the mode frequency, solid curve is the linear growth rate while the dashed curve is the saturated spectrum due to ion induced scattering.  }\label{fig:TAE_cascading}
\end{figure}

The spontaneous power  transfer from  $\delta\phi_1$ to $\delta\phi_0$ investigated above  can lead to TAE scattering to the lower frequency fluctuation  spectrum. In burning tokamak  plasma of reactor scale, where  multiple TAEs coexist, characterized by comparable frequencies and  growth rates, each TAE can interact with the   ``bath" of background TAEs, and this process can be described by an equation for spectral evolution derived from equation (\ref{eq:nl_test_eigen}). Denoting the generic test TAE with subscript $k$ and background TAE with subscript $k_1$, and summing over all background TAEs,  one obtains
\begin{eqnarray}
\hat{\epsilon}_{Ak} A_k = \sum_{k_1} \left(\Delta_0 +\chi_0\epsilon_s -\frac{\hat{C}_0}{\epsilon_s}\right)\lvert A_{k_1}\rvert^2 A_k. \label{eq:se_1}
\end{eqnarray}

Multiplying equation (\ref{eq:se_1}) with $A^*_k$, and taking the imaginary part, we then obtain the equation describing TAE nonlinear evolution due to interaction with the  bath of TAEs:
\begin{eqnarray}
\left(\partial_t-2\gamma_{L}(k)\right) I_k = \frac{2}{\partial_{\omega_k}\hat{\epsilon}_{Ak,R}} \sum_{k_1}\frac{1}{k^2_{\perp1}} \left(\frac{\hat{C}_0}{\lvert\epsilon_s\rvert^2}+\chi_0\right) \epsilon_{s,i} I_{k_1}I_k,
\end{eqnarray}
which can be rewritten as
\begin{eqnarray}
\left(\partial_t - 2\gamma_L(\omega)\right)I_{\omega} = \frac{2}{\partial_{\omega}\epsilon_{\omega,R}} \int^{\omega_M}_{\omega_L}  d \omega' V(\omega,\omega') I_{\omega'}I_{\omega},\label{eq:spectral_evolution}
\end{eqnarray}
with $I_{\omega}=\sum_k I_k\delta (\omega-\omega_k)$ being the continuum version of $I_k$, $I_k\equiv \lvert \nabla_{\perp} A_k\rvert^2$, $\omega_M$ being the highest frequency for TAE to be linearly unstable, and $\omega_L$ being the lowest frequency of TAE spectrum, which is, in fact, linearly stable, and nonlinearly excited  in the downward cascading process, as shown by Fig. \ref{fig:TAE_cascading}.  The integration kernel $V(\omega,\omega')$ is given by
\begin{eqnarray}
V(\omega,\omega')\equiv \frac{1}{k^2_{\perp \omega'}} \left(\frac{\hat{C}_0}{\lvert\epsilon_s\rvert^2} +\chi_0\right)\epsilon_{s,i}. 
\end{eqnarray}

The saturated TAE spectrum can thus be derived from the fixed point solution of equation (\ref{eq:spectral_evolution}) by taking $\partial_tI_{\omega}=0$.  The obtained integral equation, can be reduced to a differential equation noting that $I_{\omega'}$ varies in $\omega'$ much slower than $V(\omega,\omega')$, with the former varying    on the scale of $\lvert \omega_M-\omega_L\rvert \simeq \epsilon_0 \omega_A$, while the latter on the scale of $\lvert v_{it}/(qR_0)\rvert$ determined by $\epsilon_{s,I}$.  Thus, noting $I_{\omega'}=I_{\omega}-\omega_s\partial_{\omega} I_{\omega}$,     and $\lvert \omega_M-\omega_L\rvert\sim \epsilon_0\omega_A\gg\omega_s$ for the ion induced scattering process to be important as shown in Fig. \ref{fig:TAE_ISW_TAE}, one has
\begin{eqnarray}
\gamma_L(\omega)&=& - \frac{1}{\partial_{\omega}\epsilon_{\omega,R}}\int^{\omega-\omega_L}_{\omega-\omega_M} d\omega_s V(\omega_s) \left(I_{\omega}-\omega_s\partial_{\omega}I_{\omega}\right)\nonumber\\
&=&-\frac{1}{\partial_{\omega}\epsilon_{\omega,R}} \left[U_0I_{\omega}-U_1\partial_{\omega}I_{\omega}\right].\label{eq:spectrum_fixed_point}
\end{eqnarray}
with 
\begin{eqnarray}
U_0&\equiv& \int^{\omega-\omega_L}_{\omega-\omega_M} d\omega_s V(\omega_s)\simeq \int^{\infty}_{-\infty} d\omega_s V(\omega_s) \rightarrow 0,\label{eq:U_0}\\
U_1&\equiv& \int^{\omega-\omega_L}_{\omega-\omega_M} d\omega_s \omega_s V(\omega_s)\simeq \int^{\infty}_{-\infty} d\omega_s \omega_sV(\omega_s) \nonumber\\
&\simeq& \frac{\pi^{3/2}}{2 k^2_{\perp}} \left(\frac{\hat{C}_0}{\lvert \epsilon_s\rvert^2} +\chi_0\right) k^2_{\parallel s} v^2_{it}.\label{eq:U_1}
\end{eqnarray}
In deriving equations (\ref{eq:U_0}) and (\ref{eq:U_1}), it is noted that $V(\omega_s)\propto \epsilon_{s,I}$ is odd function of $\omega_s$.  Equation (\ref{eq:spectrum_fixed_point}) is the desired differential equation for the saturated spectrum, and gives
\begin{eqnarray}
I_{\omega}=\frac{2 k_{\parallel s}v_{it}\omega_M\gamma_L(\omega_M)}{U_1} - \frac{1}{U_1} \int^{\omega_M}_{\omega} \gamma_L \partial_{\omega}\epsilon_{\omega,R} d\omega,
\end{eqnarray}
which, after  integrating over the fluctuation population zone, yields the overall TAE intensity
\begin{eqnarray}
I_{Sat}\equiv \int^{\omega_M}_{\omega_L} I_{\omega} d\omega\simeq \frac{\overline{\gamma_L}}{U_1} \omega^3_T\epsilon^2_{eff},
\end{eqnarray}
with $\epsilon_{eff}\equiv 1-\omega_M/\omega_L\sim O(\epsilon)$.  Noting that $\lvert \delta B_r\rvert^2\simeq \lvert k_{\theta}\delta A_{\parallel}\rvert^2 =\lvert c k_{\theta} k_{\parallel}/(\omega k_r)\rvert^2 I_{Sat}$, one then obtains the saturation level of the magnetic fluctuations
\begin{eqnarray}
\lvert\delta B_r\rvert^2\simeq \frac{c^2\epsilon^2\epsilon^2_{eff}}{2\pi^{3/2}} \frac{\omega_T\overline{\gamma_L}k^2_r}{(\hat{C}_0/\lvert \epsilon_s\rvert^2+\chi_0)\Omega^2_{ci}\rho^2_{it}},\label{eq:TAE_em_fluctuation}
\end{eqnarray} 
which then yields, for typical parameters in burning plasma regime, the scaling law for the magnetic perturbations,
\begin{eqnarray}
\left\lvert \frac{\delta B_r}{B_0}\right\rvert^2 &\sim& \frac{m_i}{8\tau\pi^{3/2} e^2\mu_0} \frac{\overline{\gamma_L}}{\omega_T} \frac{T^2_E}{T^2_i} q^2 n^{-1}_0 \epsilon^6 R^{-2}_0\nonumber\\
&\sim& 1.2\times 10^{15} A_m q^2 n^{-1}_0\epsilon^6 R^{-2}_0 \frac{T^2_E}{T^2_i} \frac{\overline{\gamma_L}}{\omega_T}.\label{eq:TAE_fluctuation_scaling}
\end{eqnarray}
Here,  $A_m$ is the mass ratio of thermal ion to proton, and $n_0$ is the thermal plasma density. 
For typical parameters of reactors, e.g., ITER \cite{KTomabechiNF1991} or CFETR \cite{YWanNF2017}, the expected magnetic fluctuation level is $\lvert \delta B_r/B_0\rvert^2\sim O(10^{-8}\sim10^{-7})$.   It is noteworthy that  the obtained TAE magnetic perturbation  depends sensitively on the local inverse aspect ratio $\epsilon$, which is, however, not surprising,  since  TAE exist due to toroidicity ($\propto\epsilon$) induced SAW continuum gap  and 
the saturation process,  determined by ion-induced scattering,  is the TAE downward spectrum cascading (by $\sim \epsilon \omega_A$) that leads to enhanced coupling to SAW continuum.

\subsubsection{EP transport}

The TAE induced fusion alpha particle transport, can be obtained from nonlinear gyrokinetic transport theory \cite{LChenJGR1999,LChenRMPP2021}, with the expected magnetic  fluctuation level given by equation (\ref{eq:TAE_fluctuation_scaling}). Here, taking circulating EP as an example, whose transport is mainly caused by resonance overlapping induced EP orbit stochasticity \cite{TWangPoP2019}, the quasilinear transport equation for EP equilibrium distribution function evolution is \cite{LChenJGR1999,ABrizardPoP1995}
\begin{eqnarray}
\partial_t F_{0E} =-\overline{\sum_{\mathbf{k}=\mathbf{k'}+\mathbf{k''}}\Lambda^{k'}_{k''} J_{k'}\delta L_{k'}\delta H_{k''}},\label{eq:EP_transport}
\end{eqnarray}
with $\mathbf{k}=k_Z\hat{\mathbf{r}}=\mathbf{k'}+\mathbf{k''}$ denoting the bounce averaged  phase space zonal structure modulation \cite{MFalessiPoP2019} in the radial direction. Meanwhile,  the perturbed  linear EP distribution function  for well circulating EPs  can be given by \cite{ZQiuPoP2016,GFuPoFB1992}
\begin{eqnarray}
\delta H_{kE}=-\frac{e}{m}Q_k F_{0E}J_k\delta L_k\sum_{l,p}\frac{J_l(\hat{\lambda}_k) J_p(\hat{\lambda}_k) e^{-i(l-p)(\theta-\theta_{0r})}}{\omega_k-k_{\parallel}v_{\parallel}+l\omega_{tr}},\label{eq:EP_response}
\end{eqnarray}
with $QF_{0E}\equiv (\omega\partial_E -\hat{\omega}_*)F_{0E}$, $F_{0E}$ being equilibrium EP distribution function,  $\hat{\lambda}_k=k_{\perp}\hat{v}_d/\omega_{tr}$ denoting finite drift orbit width effects, and $\theta_{0r}\equiv \tan^{-1}(k_r/k_{\theta})$. Substituting equation (\ref{eq:EP_response}) into equation (\ref{eq:EP_transport}) and integrating over velocity space,  one then obtains,
\begin{eqnarray}
\partial_t n_{0E}\simeq - \partial_r \left(D_{Res} \partial_r n_{0E}\right),
\end{eqnarray}
with $n_{0E}$ being the equilibrium EP density,  the resonant circulating EP radial diffusion rate given as
\begin{eqnarray}
D_{Res}\equiv\left\langle 2\pi\sum_l \lvert\delta V_{Er,l}\rvert^2 J^2_l(\hat{\lambda}_l) \delta (\omega-k_{\parallel}v_{\parallel}+l\omega_{tr})\frac{F_{0E}}{n_{0E}}\right \rangle, \label{eq:diffusion_rate}
\end{eqnarray}
and $\lvert \delta V_{Er,l}\rvert\equiv ck_{\theta}J_k\lvert\delta\phi_k\rvert l\omega_{tr}/(B_0\omega_k)$ being the resonant EP electric-field drift velocity.  Substituting the saturated TAE fluctuation given by equation (\ref{eq:TAE_em_fluctuation}) into equation (\ref{eq:diffusion_rate}), and noting again $\lvert \delta\phi\rvert^2=\omega^2\delta B^2_r/(c^2k^2_{\theta}k^2_{\parallel})$, one obtains,
\begin{eqnarray}
D_{Res}\simeq \frac{1}{4} \frac{V_A}{k_{\parallel 0}} \left\lvert \frac{\delta B_r}{B_0}\right\rvert^2,
\end{eqnarray}
corresponding to the resonant EP transit time $\omega^{-1}_{tr,Res}$ being the wave-particle  de-correlation time. The scaling law  for TAE induced circulating EP diffusion rate can then  be derived as 
\begin{eqnarray}
D_{Res}\sim 1.3\times 10^{31} A^{1/2}_m \epsilon^6 q^3 n^{-3/2}_{0} R^{-1}_0\frac{T^2_E}{T^2_i}\frac{\overline{\gamma_L}}{\omega_T}.\label{eq:transport_scaling}
\end{eqnarray}
For typical parameters of a reactor-size tokamak,  the TAE induced resonant circulating EP  diffusion rate  can be estimated as $D_{Res}\sim 1-10 \ m^2/s$ for $\epsilon\sim 1/6-1/3$.

\subsubsection{Open questions}

The present analysis of TAE  saturation via nonlinear ion-induced scattering  extends  the previous work based on  drift kinetic theory \cite{TSHahmPRL1995}, and gaves a more quantitatively accurate estimation of the TAE saturation level and, thus, of the fusion alpha particle transport rate.  For a predictive ability of the impact on fusion plasma performance,   besides  validation of the present analytical results using first-principle-based large scale simulations,     there are several factors, which  remain to be explored.

First, the present analysis neglects  the nonuniformity of bulk  plasma, and focuses on the scattering off ion quasi mode. However,  in the nonlinear parametric decay of kinetic Alfv\'en wave (KAW), it has been demonstrated  that  bulk plasma nonuniformity may   significantly  affect    the nonlinear process, by   enhancing  the ion Compton scattering rate  by an order of magnitude  since  $\lvert \omega_{*i}\rvert \gg \lvert k_{\parallel s}v_{it}\rvert$. Furthermore, plasma nonuniformity qualitatively  breaks the parity of the decay KAW spectrum  and may have significant  implications on finite momentum transport \cite{LChenPoP2022}.  As the TAE cascading process of interest in the present review  has a one-to-one correspondence to the KAW parametric decay in slab geometry, we expect that  thermal plasma nonuniformity may also have an important consequence on the TAE saturation; this aspect  will be further explored in a separate work \cite{ZChengNF2023}.

Second, the nonlinearly generated ion quasi-mode in the present analysis, or drift sound wave as bulk plasma nonuniformity is accounted for, are both heavily ion Landau damped,. Thus, they provide a channel for nonlinearly transfer the alpha particle power to fuel deuterium-tritium ions, as originally proposed and  investigated in Ref. \cite{TSHahmPST2015} based on the   results from  Ref. \cite{TSHahmPRL1995}.  Deriving  the ion heating power from the present results and evaluating the implications to sustained burning may provide elements  of crucial importance for reactors with high temperature plasma and thus low collisonality. 

The third point to be explored is  alpha particle global   transport  and the  steady state    alpha particle  as well as bulk plasma profiles  in reactor relevant conditions. In fact, equation (\ref{eq:transport_scaling}) provides an estimate of local transport. However,   the feedbacks of alpha particle   driven instabilities on  bulk plasmas and energetic particles themselves via different channels \cite{ADiSienaNF2019,JCitrinPRL2013,LChenNF2023} should be properly taken into account.

\subsection{TAE scattering and damping by DW turbulence}
\label{sec:TAE_eDW_scattering}

The last nonlinear process to be discussed is TAE scattering by DW turbulence.  Microscopic DW  turbulence driven by expansion free energy associated with plasma nonuniformities is another significant low frequency fluctuation in magnetically confined plasmas, and is  crucial for thermal plasma transport  \cite{WHortonRMP1999}.  DWs   typically have frequencies   comparable to plasma diamagnetic frequency, and perpendicular wavelength  in the range of  thermal ion Larmor radius or even shorter, when electron dynamics plays an important role.    With different free energy sources, DWs may be driven as ion temperature gradient mode, trapped electron modes, etc., and be predominant in  different frequency range of the spectrum.  Effects of DWs on EP transport were investigated in Refs. \cite{WZhangPRL2008,ZFengPoP2013}, suggesting  that  the direct  EP transport by DWs can be negligible  due to the scale separation between EP orbit size and typical DW perpendicular wavelength. This result is consistent with theoretical expectations and should be considered a well assessed fact, despite some discussions of about a decade ago, due to the anomaly in EP confinement that was reported by AUG \cite{SGunterNF2007}, JT-60U \cite{TSuzukiNF2008}   and DIII-D  \cite{WHeidbrinkPRL2009}. In fact, it was demonstrated that the observed behavior was due to ITG driven diffusion of EP with low relative energy $E$ w.r.t. the core thermal energy $T_c$. For $E/T_c \gtrsim 10$, the EP diffusivity is typically an order of magnitude less than that of core plasma ions \cite{SZwebenNF2000}, and, thus, ``EP transport by microturbulence in reactor relevant
conditions and above the critical energy (at which plasma ions
and electrons are heated at equal rates by EPs) is negligible and
EP turbulent diffusivities have intrinsic interest mostly in present
day experiments with low characteristic values of $E/Tc$." \cite{LChenRMP2016}.  On the other hand, EP may influence the DWs stability via many mechanisms, such as thermal ion dilution \cite{GTardiniNF2007},  modification of curvature by increased pressure gradient \cite{CBourdelleNF2005}, etc. For the reference  of EP stabilization of DW turbulence, interested readers may refer to a recent review by Citrin et  al \cite{JCitrinPPCF2023}.

With  two fundamental  fluctuations groups coexisting, DWs and SAWs, or more precisely drift Alfv\'en waves (DAWs), characterized by distinct spatial and temporal scales,   and dominating  transport in  different energy ranges, it is natural to consider their mutual effects. The nonlinear interactions of DWs and SAW instabilities via   ZFS have been proposed theoretically  \cite{LChenRMP2016, LChenPRL2012,ZQiuPoP2016,LChenNF2001,ZQiuNF2016,FZoncaPPCF2015} and investigated numerically \cite{YTodoNF2010,DSpongPoP1994,HZhangPST2013}, and  were suggested as possible interpretation of  experimental observation of confinement improvement  with large fraction of  EPs \cite{JCitrinPRL2013,ADiSienaNF2019,ADiSienaJPP2021,SMazziNP2022}. This ``indirect channel"  remains to be investigated in more detail due to the multiple facets consiting of complex nonlinear behaviours.    It was proposed, in our  recent work, that the DWs and SAW instabilities  can also mutually  interact  via direct nonlinear mode coupling processes,  which can lead to, e.g.,  suppression of TAE due to the scattering by  finite amplitude  electron DW (eDW)   \cite{LChenNF2022}.  The ``inverse" process, on the other hand, shows that finite amplitude  TAE has negligible effects on the eDW stability \cite{LChenNF2023}.
This paradigm,  proposed using  TAE and eDW as example,   can be generalized to include other effects such as   trapped electron contribution.  Here, we will briefly review the TAE scattering by finite amplitude eDW.

\begin{figure}
\includegraphics[scale=0.60]{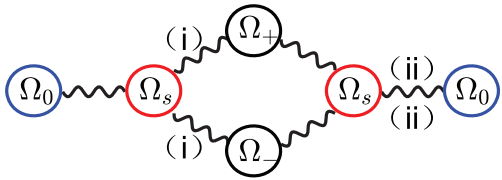} 
\caption{Cartoon of  the two-step nonlinear  process of TAE scattering by eDW. The first process corresponds to short scale KAW sidebands generation due to eDW scattering, while the second corresponds to their feedback on  the test TAE. }\label{fig:TAE_eDW_scattering}
\end{figure}

\begin{figure}
\includegraphics[scale=0.60]{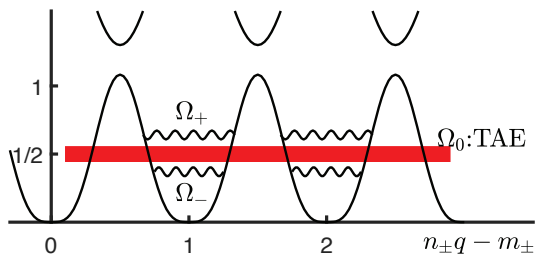} 
\caption{Cartoon of  upper and lower   KAWs generation   due to TAE-eDW scattering, and coupling to continuum.}\label{fig:KAW_generation}
\end{figure}

The TAE-eDW scattering process, can be understood as the test TAE ``linear" stability in the presence of finite amplitude eDW, and can be considered as a two-step process. In  the first process, short wavelength upper and lower kinetic Alfv\'en wave (KAW) sidebands are generated, with frequency comparable to TAE  and  high toroidal mode number determined by  eDW. In the second step, KAW     then couple with eDW and feed back on   the  test TAE,  modifying its dispersive properties and stability as shown in Fig. \ref{fig:TAE_eDW_scattering}.  The damping of the mode-converted upper and lower KAWs, as shown in Fig. \ref{fig:KAW_generation}, then lead to the damping of the test TAE.  

\subsubsection{KAW generation}

We start from the upper sideband $\Omega_+$ generation channel due to test TAE $\Omega_0$  and eDW $\Omega_s$ coupling,   noting that the analysis for $\Omega_-$ is similar. The linear and nonlinear particle responses to $\Omega_+$, can be derived noting the $\lvert k_{\parallel}v_{te}\rvert\gg\lvert\omega_+\rvert\gg \lvert k_{\parallel}v_{ti}\rvert$ ordering,  and one has, to the leading order,
\begin{eqnarray}
\delta H^{(1)}_{+i}&\simeq& \frac{e}{T_i}F_M \left(1-\frac{\omega_{*i}}{\omega}\right)_+ J_+\delta\phi_+,\\
\delta H^{(1)}_{+e}&\simeq& -\frac{e}{T_e} F_M \left(1-\frac{\omega_{*e}}{\omega}\right)_+\delta\psi_+.
\end{eqnarray}

The nonlinear ion response  to $\Omega_+$ can be derived as
\begin{eqnarray}
\delta H^{(2)}_{+,i}\simeq -i\frac{\Lambda^s_0}{2\omega_0} J_0J_s \frac{e}{T_i}F_M \left(\frac{\omega_{*i}}{\omega}\right)_s\delta\phi_s\delta\phi_0,
\end{eqnarray}
having noted  the linear ion response to $\Omega_0$ and $\Omega_s$.  On the other hand, nonlinear electron contribution to upper KAW  can be neglected, since $\Omega_s$ is predominantly electrostatic. Substituting the particle responses into quasi-neutrality condition, we then have,
\begin{eqnarray}
\delta\psi_+=\sigma_{*+}\delta\phi_+ + i\frac{\Lambda^s_0}{2\omega_0} D_+ \delta\phi_0\delta\phi_s, \label{eq:UKAW_qn}
\end{eqnarray}
where $\sigma_{*k}$, derived in equation (\ref{eq:sigma_k_expression}),  denotes the deviation from ideal MHD condition due to plasma nonuniformity and/or FLR effects, while $D_+=\tau(\omega_{*i}/\omega)_sF_+/(1-\omega_{*e}/\omega)_+$ denotes nonlinear contribution with $F_+=\langle J_0J_sJ_+F_{M}/n_0\rangle_v$.   The other equation for $\Omega_+$, can be derived from nonlinear vorticity equation, by substituting the linear particle responses to $\Omega_0$ and $\Omega_s$ into the Reynolds stress term
\begin{eqnarray}
&&\tau b_+ \left[ \left(1-\frac{\omega_{*i}}{\omega}\right)_+\frac{(1-\Gamma_+)}{b_+} \delta\phi_+ -\left(\frac{V^2_A}{b}\frac{k_{\parallel} b k_{\parallel}}{\omega^2}\right)_+\delta\psi_+     \right]\nonumber\\
&=& -i\frac{\Lambda^s_0}{2\omega_0} \gamma_+\delta\phi_0\delta\phi_s,\label{eq:UKAW_vorti}
\end{eqnarray} 
with $\gamma_+ = \tau [\Gamma_s-\Gamma_0+(\omega_{*i}/\omega)_s(F_+-\Gamma_s)]$.

Combining equations (\ref{eq:UKAW_qn}) and (\ref{eq:UKAW_vorti}), one obtains, the equation for upper KAW generation due to $\Omega_0$ and $\Omega_s$ coupling
\begin{eqnarray}
\tau b_+\epsilon_{A+}\delta\phi_+=-i(\Lambda^s_0/2\omega_0) \beta_+ \delta\phi_s\delta\phi_0,\label{eq:dr_kaw_u}
\end{eqnarray}
where
$\epsilon_{A+}$
is the linear SAW/KAW operator given by equation (\ref{eq:SAW_operator}) with curvature coupling term neglected due to high TAE frequency range, and
\begin{eqnarray}
\beta_+&=&\tau(\Gamma_s-\Gamma_0)
 +\tau \left(\frac{\omega_{*i}}{\omega}\right)_s \left[F_+-\Gamma_s- \left(\frac{k_{\parallel}bk_{\parallel}}{\omega^2}\right)_+ \frac{\tau V^2_A F_+}{(1-\omega_{*e}/\omega)_+} \right].
\end{eqnarray}

The generation of lower KAW $\Omega_-$ due to $\Omega^*_0$ and $\Omega_s$ coupling, can be derived similarly as
\begin{eqnarray}
\tau b_-\epsilon_{A-}\delta\phi_- =  i(\Lambda^s_0/2\omega_-)\beta_-\delta\phi_s\delta\phi^*_0, \label{eq:dr_kaw_l}
\end{eqnarray}
with
\begin{eqnarray}
\beta_-=\tau(\Gamma_s-\Gamma_0)+\tau\left(\frac{\omega_{*i}}{\omega}\right)_s\left[F_--\Gamma_s-\left(\frac{k_{\parallel}bk_{\parallel}}{\omega^2}\right)_- \frac{\tau V^2_AF_-}{(1-\omega_{*e}/\omega)_-}\right].
\end{eqnarray}

\subsubsection{Feedback to $\Omega_0$ and consequence on TAE stability}

The effect of eDW scattering on the test TAE, can be derived by  accounting for  feedback of  $\Omega_\pm$  via nonlinear coupling to $\Omega_s$. Here, we only discuss the   contribution to $\Omega_0$ by nonlinear coupling between $\Omega_+$ and $\Omega^*_s$, noting that the contribution due to $\Omega_-$ and $\Omega_s$ coupling can be derived similarly.

The nonlinear ion response to $\Omega_0$ can be derived as
\begin{eqnarray}
\delta H^{(2)}_{0i}&\simeq& \frac{e}{T_i} F_{M}\left( \frac{\omega_{*i}}{\omega}\right)_s\left[ i\frac{\Lambda^s_0}{2\omega_0} J_sJ_+\delta\phi^*_s\delta\phi_+ +\left(\frac{\Lambda^s_0}{2\omega_0}\right)^2 J_0J^2_s\lvert\delta\phi_s\rvert^2\delta\phi_0  \right]\nonumber\\
&+&    \delta\phi_-\  \mbox{contribution},
\end{eqnarray}
with the second term corresponding to    $\delta H^{(2)}_{+i}$ contribution. The nonlinear electron response to $\Omega_0$ is negligible. The quasi-neutrality condition then yields
\begin{eqnarray}
\delta\psi_0=\left(\sigma_{*0}+\alpha_0\lvert\delta\phi_s\rvert^2\right)\delta\phi_0 - i(\Lambda^s_0/2\omega_0) D^+_0 \delta\phi^*_s\delta\phi_+ + \delta\phi_-\  \mbox{contribution}, \label{eq:TAE_test_qn}
\end{eqnarray}
with $\alpha_0=-(\Lambda^s_0/2\omega_0)^2\tau(\omega_{*i}/\omega)_sF_2$, $F_2\equiv \langle J^2_0J^2_sF_{M}/n_0\rangle_v$ mainly contributing to nonlinear frequency shift, while $D^+_0=\tau(\omega_{*i}/\omega)_s F_+/(1-\omega_{*e}/\omega)_0$.

The other equation for $\Omega_0$ can then be derived from nonlinear vorticity equation, as
\begin{eqnarray}
&&\tau b_0 \left\{\left[\left(1-\frac{\omega_{*i}}{\omega_0}\right)_0 \frac{(1-\Gamma_0)}{b_0} +\alpha^+_0\lvert \delta\phi_s\rvert^2\right]\delta\phi_0-\left(\frac{V^2_A}{b} \frac{k_{\parallel} b k_{\parallel}}{\omega^2}\right)_0\delta\psi_0\right\}\nonumber\\
& =& i\frac{\Lambda^s_0}{2\omega_0}\gamma^+_0\delta\psi^*_s\delta\phi_+  + \delta\phi_-\  \mbox{contribution}.\label{eq:TAE_test_vorti}
\end{eqnarray}

Substituting equation (\ref{eq:TAE_test_qn}) into (\ref{eq:TAE_test_vorti}),  and neglecting the nonlinear frequency shift while focusing on the stability of the test TAE due to scattering by background eDW,     one then obtains 
\begin{eqnarray}
\tau b_0\epsilon_{A0} \delta\phi_0=i\frac{\Lambda^s_0}{2\omega_0} \beta^+_0\delta\phi^*_s\delta\phi_+ + \delta\phi_-\  \mbox{contribution}. \label{eq:TAE_test_nl}
\end{eqnarray}

\begin{figure}
\includegraphics[scale=0.20]{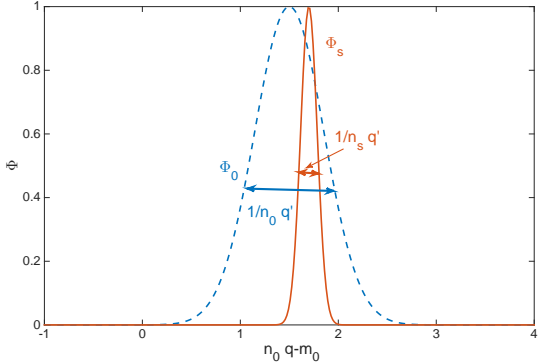} 
\caption{Cartoon of scale separation between TAE and eDW, with the dashed curve being the sketched parallel mode structure of a TAE poloidal harmonic, while the solid curve being the parallel mode structure of eDW with much smaller radial width than that of the TAE.}\label{fig:scale_separation}
\end{figure}

Substituting $\delta\phi_+$ from equation (\ref{eq:dr_kaw_u}) into (\ref{eq:TAE_test_nl}),  one obtains,
\begin{eqnarray}
\tau b_0 \epsilon_{A0}\delta\phi_0=\left[\left(\frac{\Lambda^s_0}{2\omega_0}\right)^2 \beta^+_0\delta\phi^*_s \frac{\beta_+}{\tau b_+\epsilon_{A+}}\delta\phi_s \right]\delta\phi_0 + \delta\phi_-\  \mbox{contribution}, \label{eq:TAE_test_nl_1}
\end{eqnarray}
which can be solved noting the spatial  scale separation between $\delta\phi_0$ and $\delta\phi_s$, as sketched  in Fig. \ref{fig:scale_separation}.  
Thus, the nonlinear coupling processes occur  in  in a narrow region of the eDW localization.
   Expanding $\delta\phi_0=\Phi_0(\mathbf{x}_0) +\tilde{\Phi}_0(\mathbf{x}_s,\mathbf{x}_0)$ with $\mathbf{x}_0=(R/n_0, r/m_0, 1/n_0q')$, $\mathbf{x}_s=(R/n_s, r/m_s, 1/n_sq')$ and $\lvert\tilde{\Phi}_0\rvert/\lvert\Phi_0\rvert\sim O(\lvert e\delta\phi_s/T_e\rvert^2)\ll1$,  equation (\ref{eq:TAE_test_nl_1}) becomes, after averaging over $\mathbf{x}_s$ scale,
\begin{eqnarray}
 \tau b_0 \epsilon_{A0}\Phi_0=\left\langle\left(\frac{\Lambda^s_0}{2\omega_0}\right)^2 \beta^+_0\delta\phi^*_s \frac{\beta_+}{\tau b_+\epsilon_{A+}}\delta\phi_s \right\rangle_s\Phi_0 + \delta\phi_-\  \mbox{contribution}, \label{eq:TAE_test_nl_2}  
\end{eqnarray}
with  $\langle\cdots\rangle_s$ denoting averaging over eDW scales  
\begin{eqnarray}
\langle (\cdots)\lvert\delta\phi_s\rvert_s\rangle_s \equiv \lvert A_{n_s}\rvert^2\left(\int^{\infty}_{\infty} dz_s \lvert\Phi_s(z_s)\rvert^2\right)^{-1}\int^{\infty}_{\infty} dz_s(\cdots)\lvert\Phi_s(z_s)\rvert^2.
\end{eqnarray}

Equation (\ref{eq:TAE_test_nl_2}) can then be solved noting that the absorption due to    $\mbox{Im}(1/\epsilon_{A+})$ can be expressed as $\mbox{Im}(1/\epsilon_{A+})=-\pi\delta(\epsilon_{A+})\simeq -(\pi/4\sigma_{*+})\delta (z^2_s-z^2_+)$ with $z^2_+=(1-\omega_{*i}/\omega)_+(1-\Gamma_+)(\omega/\omega_A)^2_+/(b_+\sigma_{*+})$. This implies that  KAW are absorbed locally. Thus, expanding  $\beta_+$  with respect to $b_s$ noting the two spatial scale separation $k^2_{\perp+}\simeq k^2_{\perp s}+2k_{rs}k_{r0}$, and properly reinstating the lower KAW contribution, 
one  obtains,
\begin{eqnarray}
\tau b_0\left[\epsilon_{A0}+i\nu(k_{r0}\rho_i)^2\right]\Phi_0 = 0, \label{eq:TAE_test_nl_3}
\end{eqnarray}
with $\nu=\nu_++\nu_-$, and
\begin{eqnarray}
\nu_{\pm}\simeq \pi\left(\frac{\Omega_{ci}}{\omega_0}\right)^2 \sum_{n_s} \lvert A_{n_s}\rvert^2 \left[\left(\tau+\frac{\sigma_s}{2\Gamma_s}\right)\frac{\partial\Gamma_s}{\partial b_{\theta s}} \right]^2 \frac{b_{\theta s}\hat{s}^2}{\sigma^2_{\pm s}z_{\pm}} \left\lvert \frac{\partial\Phi_s}{\partial z_s}\right\rvert^2_{z_{\pm}}. 
\end{eqnarray}

Equation (\ref{eq:TAE_test_nl_3}) can be solved perturbatively in ballooning space, $\eta$.  Letting $\hat{\Phi}_0(\eta)$ being the lowest order linear  eigenmode satisfying $\hat{b}_0\hat{\epsilon}_{A0}(\eta, \partial_{\eta},\omega_0)\hat{\Phi}_0(\eta)=0$, and expanding $\omega_0=\omega_{0,R}+i\gamma_{AD}$ with $\gamma_{AD}$ being the eDW scattering induced test TAE damping rate, equation (\ref{eq:TAE_test_nl_3}) then gives, 
\begin{eqnarray}
\frac{2\gamma_{AD}}{\omega_{0,R}}\left\langle \hat{\Phi}_0\hat{b}_0\hat{\Phi}_0\right\rangle_{\eta} =-\left\langle \hat{\Phi}_0\hat{b}_0\nu b_{\theta0}\hat{s}^2\eta^2\hat{\Phi}_0\right\rangle_{\eta},
\end{eqnarray}
with $\langle\cdots\rangle_{\eta}\equiv\int^{\infty}_{-\infty}(\cdots) d\eta$.   Noting equation (\ref{eq:TAE_inertial_test}) for TAE, we obtain
\begin{eqnarray}
\frac{\gamma_{AD}}{\omega_{0,R}}=-\frac{1}{4}\frac{\nu b_{\theta0}\hat{s}^2}{\sqrt{-\Gamma_l\Gamma_u}} \sim O(10^{-2} - 10^{-1}),
\end{eqnarray}
as estimated using  typical parameters, i.e., $\lvert \Omega_{ci}/\omega_{0,R}\rvert\sim O(10^2)$, $\sum_{n_s}\lvert A_{n_s}\rvert^2\sim \lvert e\delta\phi_s/T_e\rvert^2\sim O(10^{-4})$, $b_{\theta s}\sim\hat{s}\sim\tau\sim O(1)$ and $4\sqrt{-\Gamma_l\Gamma_u}\sim O(\epsilon^2)\sim O(10^{-2}-10^{-1})$.  The eDW scattering induced TAE damping rate  is comparable to the TAE growth rate due to EP drive \cite{LChenRMP2016}, and can significantly reduce or completely suppress TAE fluctuations with sufficiently large eDW intensity.  This may imply improved fusion alpha particle confinement  in the existence of micro turbulence, and consequently, enhanced thermal plasma heating.

As the nonlinearly generated KAW quasi-modes are dissipated by  predominantly electron Landau damping \cite{AHasegawaPoF1976,LChenRMPP2021},  the resulting electron heating rate can be estimated as
\begin{eqnarray}
\left(\frac{d\beta_e}{dt}\right)_{AD} = 4\lvert\gamma_{AD}\rvert\left\lvert\frac{\delta B_{\perp}}{B_0}\right\rvert^2\simeq O(10^{-2}-10^{-1})  s^{-1}, 
\end{eqnarray}
which, for typical parameters,  can be comparable to the electron heating by alpha particle slowing down, and potentially, contribute significantly to the ``anomalous" electron heating in burning plasmas \cite{LChenNF2022}. 

The present analysis, using TAE scattering by ambient  eDW as an example to demonstrate the novel physics of direct cross-scale interaction among meso/macro-scale SAW instabilities and micro-scale DW turbulence, and the obtained results are expected to be, at least  qualitatively,  applicable to other Alfv\'en eigenmodes,  such as reversed shear Alfv\'en eigenmode (RSAE), interacting with various branches of DWs, including the physics of finite temperature gradients or  trapped electrons. These applications to more realistic scenarios can be investigated for a more detailed understanding of the SAW  stabilities  and   fusion alpha particle confinement  in reactors.

\section{Summary} \label{sec:summary}

Using  toroidal Alfv\'en eigenmode (TAE) nonlinear saturation  due to  mode-mode coupling as example, we show that, nonlinear gyrokinetic theory is not only efficient, but  also  necessary  to investigate various  crucial physics in the nonlinear mode coupling processes of SAW instabilities.  This necessity occurs since SAW instabilities often have    a small scale structure associated with the SAW continuum related to equilibrium magnetic geometry  and plasma nonuniformity of magnetically confined  fusion devices. The nonlinear coupling is,  thus, dominated through  perpendicular scattering \cite{LChenEPL2011}. Three main processes developed in the past decade  are briefly reviewed, i.e., the zonal field structure generation by TAE \cite{LChenPRL2012}, TAE spectral cascading due to ion induced scattering \cite{TSHahmPRL1995,ZQiuNF2019a}, and cross-scale interaction with electron drift wave (eDW)   via  direct nonlinear interaction \cite{LChenNF2022}.  The fundamental physics involved in the three processes are reviewed in a pedagogical way,  discussing parameter regimes for them to occur and dominate,  and state-of-art developments as well as open questions are   introduced.  These understandings present a road map for   a comprehensive and quantitative study of SAW instability spectrum in fusion reactors, and provide guidance for large scale simulations using realistic geometry and plasma parameters.

It is obvious that   the  nonlinear mode coupling processes reviewed in the present work,   
 and the  self-consistent EP transport should be considered on the same footing for the comprehensive understanding nonlinear dynamics and self-organization in burning fusion plasmas.  There exists a general theoretical framework that addresses the former wave-wave coupling processes, which can be described by the nonlinear radial envelope equation in the form of a  nonlinear  Schr$\ddot{o}$dinger  equation. Meanwhile,  the latter nonlinear wave-particle interactions and ensuing EP transport may be  described by  the Dyson-Schr$\ddot{o}$dinger model,  
   providing  a general theoretical framework for SAW nonlinear dynamics and EP transport in burning plasmas \cite{FZoncaNJP2015,LChenRMP2016,FZoncaJPCS2021,MFalessiPoP2019,MFalessiNJP2023}.

\section*{Acknowledgement}
This work was  supported by  the National Science Foundation of China under Grant Nos. 12275236 and 12261131622, Italian Ministry for Foreign Affairs and International Cooperation Project under Grant  No. CN23GR02, and ``Users of Excellence program of Hefei Science Center CAS" under Contract No. 2021HSC-UE016.
 This work was was supported by the EUROfusion Consortium, funded by the European Union via the Euratom Research and Training Programme (Grant Agreement No. 101052200 EUROfusion). The views and opinions expressed are, however, those of the author(s) only and do not necessarily reflect those of the European Union or the European Commission. Neither the European Union nor the European Commission can be held responsible for them.

\section*{Declarations}
\begin{itemize}
\item  Conflict of interest 

The authors have no conflicts to disclose. 

\item Data availability

The data that support the findings of this study are available from the corresponding author upon reasonable request.

\end{itemize}


\end{document}